**Growing Mars fast: High-resolution GPU simulations of embryo formation**

J. M. Y. Woo[a], S. Grimm[b], R. Brasser[c], J. Stadel[a]

[a]Institute for Computational Science, University of Zürich, Winterthurerstrasse 190, 8057 Zürich, Switzerland

[b]Center for Space and Habitability, University of Bern, Gesellschaftsstrasse 6, 3012 Bern, Switzerland

[c]Earth Life Science Institute, Tokyo Institute of Technology, Meguro-ku, Tokyo 152-8550, Japan

## Abstract

Recent high precision meteoritic data improve constraints on the formation timescale and bulk composition of the terrestrial planets. High resolution N-body simulations allow direct comparison of embryo growth timescale and accretion zones to these constraints. In this paper, we present results of high resolution simulations for embryo formation from a disc of up to 41,000 fully-self gravitating planetesimals with the GPU-based N-body code *GENGA*. Our results indicate that the growth of embryos are highly dependent on the initial conditions. More massive initial planetesimals, a shorter gas disc decay timescale and initially eccentric Jupiter and Saturn (EJS) all lead to faster growth of embryos. Asteroid belt material can thereby be implanted into the terrestrial planet region via sweeping secular resonances. This could possibly explain the rapid growth of Mars within 10 Myr inferred from its Hf-W chronology. The sweeping secular resonance almost completely clears the asteroid belt and deposits this material in the Mercury-Venus region, altering the composition of embryos there. This could result in embryos in the Mercury-Venus region accreting an unexpectedly high mass fraction from beyond 2 AU. Changing the initial orbits of Jupiter and Saturn to more circular (CJS) or assuming embryos formed in a gas free environment removes the sweeping secular resonance effect and thus greatly decreases material accreted from beyond 2 AU for Mercury-Venus region embryos. We therefore propose that rock samples from Mercury and Venus could aid greatly in deducing the condition and lifetime of the initial protoplanetary gas disc during planetesimal and embryo formation, as well as the initial architecture of the giant planets.

## 1. Introduction

Our understanding of terrestrial planet formation has dramatically improved over the last two decades, thanks in part to the development of increasingly-powerful computational hardware and sophisticated software, both of which allow faster and more sophisticated numerical calculation for planet formation processes. At the same time, improved elemental and isotopic measurements of samples from Earth and the other solar system objects, such as meteorites from the Moon, Mars and the asteroid 4 Vesta grants us the opportunity to further constrain the bulk composition and the formation timescale of the samples' parent bodies. Combining high resolution N-body simulations and the meteoritic data is an effective way to tackle the problem of early Solar System formation. For instance, Mars' rapid formation currently still remains a puzzle. Simulating the growth of Mars from the pre-runaway stage requires N-body simulations with much higher resolution than was possible previously (Morishima et al. 2013).

## 1.1. Late stage of planet formation - from planetesimals to the terrestrial planets

Planetesimals are thought to form in a circumstellar disc via coagulation, aggregation and compaction (Dominik et al., 2007; Lissauer, 1993; Wetherill, 1980) of dust grains or/and by streaming instability, which is a process involving gravitational collapse of concentrated dust grains (Johansen et al., 2007, 2015). The streaming instability helps overcoming the meter-sized sticking and inward drifting barriers (Benz 2000; Weidenschilling 1977) and suggests that the first planetesimals form directly via gravitational collapse of sub-meter size objects, reaching ~100 km in diameter, without passing through the m- to km-sized regime (Cuzzi et al. 2008; Johansen et al. 2007, 2015). This scenario of forming the initial planetesimals "big" has also been suggested by Morbidelli et al. (2009) and is supported by the current asteroid belt's size-frequency distribution (SFD) (Delbo et al., 2017; 2019).

After tens to hundreds of km-sized planetesimals formed in the disc, the dynamical evolution of the system would be dominated by gravitational interaction between planetesimals. The formation of planetesimals could be within 1 Myr of the birth of the Solar system according to the ages of various iron meteorites' parent bodies (Kruijer et al., 2014, 2017a) and thus planetesimal formation likely occurred within the Sun's primordial gaseous disc (Armitage et al., 2003; Haisch et al., 2001; Strom et al., 1989). The subsequent growth of planetesimals (in the 10s-100s km size range) to embryos with radii > 1000 km could also be affected by the existence of the gaseous disc.

Since the dynamical evolution of the system is now dominated by gravity, numerical N-body simulations, solving the Hamiltonian (the sum of kinetic and gravitational potential energy) of the system, are adopted to study terrestrial planet's formation from planetesimals to fully formed planets. Since the early 90's, several different symplectic N-body integrators have been developed (Chambers, 1999; Duncan et al., 1998; Grimm and Stadel, 2014; Stadel, 2001; Wisdom and Holman, 1991), which aid us in understanding the dynamical history and stability of a planetary system. However, simulating the growth of the terrestrial planets beginning with planetesimals remains challenging because of the large $N$ problem: the computational time generally scales as $N^2$, where $N$ is the particle number of a system. In addition, the high number of required integration time-steps as well as special treatment required for bodies within close encounters further increases the computational challenge of such simulations. Therefore most previous studies of the terrestrial planets formation (e.g. Brasser et al., 2016; Chambers, 2001; Clement et al., 2018, 2019; Jacobson & Morbidelli, 2014; O'Brien et al., 2006; Raymond et al., 2006, 2009; Walsh et al., 2011; Woo et al., 2018) significantly simplify the disc and approximate it with just tens of embryos and hundreds to thousands of planetesimals in order to achieve the necessarily long integration times of >100 Myr on desktop or server CPU cores.

Based on a series of N-body simulation studies with dedicated hardware performed by Kokubo & Ida (1996, 1998, 2000, 2002), the growth of embryos from planetesimals can be divided into two stages: the runaway growth and the oligarchic growth. During the runaway growth, the larger embryos grow faster than the smaller embryos (Kokubo and Ida, 1996) and hence the mass ratio between two bodies increases with time (Kokubo and Ida, 2000). When the massive embryo grows to a certain size, it starts to scatter the surrounding planetesimals, leading to an increase in the system's random velocity and hence a slower growth rate of the larger embryo (Ida and Makino, 1993). The system thus enters the phase of oligarchic growth in which the larger embryo grows slower than the smaller one and the mass ratio between two embryos decreases with time (Kokubo and Ida, 1998, 2000), leading to a bimodal size distribution. During the oligarchic growth phase the spacing between the embryos is about 5 to 10 mutual Hill radii due to orbital repulsion (Kokubo and Ida, 1998, 2000). Then it is succeeded by a chaotic phase and the more massive terrestrial planets formed from giant impacts between embryos lasting for hundreds million years (e.g. Agnor et al., 1999; Chambers, 2001).

### 1.2. Numerical approach in studying the late stage of the terrestrial planet formation

Applying a direct N-body approach to simulate the formation of terrestrial planets from pre-runaway planetesimals remains difficult even today. Even though there is a recent study pushing the resolution to $10^6$ particles (Wallace and Quinn, 2019), simulating the growth of the terrestrial planets from such a high *N* system requires an extremely long time (more than a year, Clement et al. (2020)) on modern server nodes. Most up to date studies involving N-body methods that run within a reasonable time on fast desktop computers can only form the terrestrial planets from embryos after the initial runaway and oligarchic growth phases. On top of the bimodal embryo-planetesimal size-frequency distribution obtained by early studies (Kokubo and Ida, 1998, 1995), the initial conditions of most terrestrial planet simulations consist of tens to hundreds of relatively large embryos embedded in a swarm of thousands of smaller planetesimals within the orbit of Jupiter (e.g. Brasser et al., 2016; Clement et al., 2018, 2019; Jacobson and Morbidelli, 2014; Lykawka and Ito, 2019; O'Brien et al., 2006; Raymond et al., 2006, 2009; Walsh et al., 2011). The masses of the embryos are assumed either to be all equal or calculated by a semi-analytic theory based on the oligarchic growth model (Chambers, 2006). In these simulations the planetesimals feel the gravitational force of the embryos and the damping forces of the disc, but not their mutual gravity, so that the $N^2$ scaling of the force calculation only applies to the embryos. Although starting the simulation with fully formed embryos saves orders of magnitude of computational time on regular CPUs, this approach loses track of some crucial early features of terrestrial planet formation during planetesimal accretion. This phase is crucial to understand the formation of Mars.

Studying the rapid formation of Mars (< 10 Myr, Dauphas and Pourmand, 2011; Tang and Dauphas, 2014; cf. Marchi et al., 2020) becomes difficult if we begin the simulations with lunar- to Mars-sized embryos because Mars is often already half or even fully formed. The bulk composition of Mars would also be dominated by the initial composition of only a few embryos, which leads to large uncertainties in estimating Mars' bulk composition, even when averaged over hundreds of simulations (Woo et al., 2018). Hence, high resolution simulations of terrestrial planet formation beginning from only planetesimals is warranted, both from a computational aspect and a cosmochemical one if we intend to combine meteoritic data with N-body results in a more accurate way.

Earlier studies forming the terrestrial planets from planetesimals invoked a hybrid N-body code. When planetesimals in the disc are still small (typically with radius $r < 100$ km), some hybrid codes adopt a statistical approach to predict the mass and dynamical evolution of planetesimals (Kenyon and Bromley, 2006; Leinhardt et al., 2009; Levison et al., 2012; Morishima, 2015; Walsh and Levison, 2019). After an object in the disc reaches a threshold radius the code promotes it to the N-body regime in which gravitational forces between these massive bodies are solved (Kenyon and Bromley, 2006; Levison et al., 2012; Morishima, 2015). Recently, Walsh & Levison (2019) studied the terrestrial planet formation from planetesimals as small as $r$ = 30 km. They applied the "*LIPAD*" (Levison et al., 2012) hybrid N-body code based on the symplectic integrator SyMBA (Duncan et al., 1998) that includes collision fragmentation algorithms based on the works of Benz & Asphaug (1999) and Morbidelli et al. (2009). Their results indicate that oligarchic growth was never achieved in the outer and inner regions of the inner solar system at the same time: in some cases the inner disc has already entered the giant impact phase but the outer disc is still in the runaway growth stage. Walsh & Levison (2019) attribute this lack of growth in the asteroid belt region due to collisional grinding between planetesimals.

The emergence of GPU-based N-body codes poses a novel way to study systems with a high number of particles. The GPU accelerated N-body integrator "*GENGA*" (Grimm and Stadel, 2014) takes advantage of the large number of computing cores in GPU cards which can perform the same instructions on multiple threads in parallel. Grimm and Stadel (2014) were able to speed up both the mutual force calculation as well as the close encounters. The speed increase over a regular CPU core for high $N$ is at least a factor of 30 (Grimm and Stadel, 2014). Recent studies have taken this advantage to study the formation of the terrestrial planets with *GENGA* (Clement et al., 2020; Hoffmann et al., 2017). Beginning a simulation with 2000 equal mass planetesimals in between the terrestrial planet region and the asteroid belt region, Hoffmann et al. (2017) found that the mass-distance distribution of the final system is controlled by the orbits of the giant planets, with giant planets on eccentric orbits producing more massive terrestrial planets in tighter orbits than when the giants reside on circular orbits. They further suggest that some exoplanetary systems, e.g. CoRoT-7, HD 20003 and HD 20781 may host undetected giant planets based on their simulation results. Another published study on the terrestrial planet formation from planetesimals by *GENGA* was performed by Clement et al. (2020). They first formed embryos in several narrow annuli from small planetesimals with $r$ = 100 km and then combined all annulus simulations into one terrestrial planet forming simulation. The annuli did not connect to each other so that specific segments of the disc were left unexplored. Clement et al. (2020) found that compared to other embryo forming simulations (Carter et al., 2015; Walsh and Levison, 2019) their simulations formed fewer embryos in the Venus and Earth region, which may suggest Venus and Earth suffered fewer giant impacts than previously expected. Their full simulation took nearly two years to complete on *NVIDIA Tesla GK110 (K20X)* "Kepler" accerators, which demonstrates that extremely high resolution N-body simulations for terrestrial planet formation are still greatly limited by the computation power available today (Portegies Zwart, 2020).

## 1.3. Diverse models for the terrestrial planet formation

Most terrestrial planet formation N-body simulations begin at the post-runaway stage focuses on the "classical" case, in which the solid surface density of the protosolar disc follows the "Minimum Mass Solar Nebula" (MMSN) (Hayashi, 1981) with Jupiter and Saturn residing close to their current location without migration (e.g. Chambers 2001; Raymond et al. 2009). Recently Hoffman et al. (2017), Walsh and Levison (2019) and Clement et al. (2020) have studied the problem from the pre-runaway stage. However, a recurring problem of the classical model is the "massive Mars" problem, in that the simulations often form a Mars analogue that is more than a few times as massive as its current mass (e.g. Chambers, 2001; O'Brien et al., 2006; Raymond et al., 2009; Wetherill, 1980). This problem still persists in high resolution simulations, regardless of whether fragmentation is included (Clement et al., 2020; Quintana et al., 2016; Walsh and Levison, 2019). Two approaches have been proposed to solve this problem.

The first approach is to rely on the gravitational influence from the giant planets to clear the region near Mars' orbit and the asteroid belt region. The "Grand Tack" scenario suggested that Jupiter and Saturn could have migrated inward through the asteroid belt then back to their current locations when the gas disc still exists (Walsh et al., 2011). This process truncates the inner disc to within 1 AU (Brasser et al., 2016; Jacobson and Morbidelli, 2014; Walsh et al., 2011; Walsh and Levison, 2019). The region of Mars to the asteroid belt can also be cleared after the gas disc dissipated by, for example, assuming initially more excited orbits of Jupiter and Saturn (Lykawka and Ito, 2019; Raymond et al., 2009), sweeping secular or mean-motion resonance (Bromley and Kenyon, 2017; Lykawka and Ito, 2013; Nagasawa et al., 2000, 2005) or by the "Nice model" giant planet instability (Tsiganis et al., 2005) which might have occurred just after 1 to 100 Myr of the dispersal of gas disc (Clement et al., 2018, 2019; Costa et al. 2020; Mojzsis et al., 2019; Ribeiro et al., 2020).

Another approach is to assume a specific initial solid surface density profile that concentrates the majority of the disc mass < 1 AU. Three models have been proposed: (1) the "annulus" model in which all solids forming the terrestrial planets concentrate in an annulus between 0.7 AU and 1 AU (Hansen, 2009); (2) the "depleted disc" model where the solid surface density in the Mars to asteroid belt region drops compared to the MMSN (Izidoro et al., 2014; Mah and Brasser, 2021) and (3) the "empty asteroid belt" model in which a much steeper solid surface density function than the MMSN is assumed (Izidoro et al., 2015; Raymond and Izidoro, 2017).

Finally, the pebble accretion model, which was originally proposed for explaining the rapid core formation of the giant planets by accreting mm to cm sized inward drifting objects (Lambrechts and Johansen, 2012), has also been shown to successfully reproduce the low mass of Mars and the asteroid belt (Levison et al., 2015b).

### 1.4. Meteorites - indicators of planet formation processes and timescales

With all the previous models claiming a fairly high probability of forming a low mass Mars and a nearly empty asteroid belt based only on dynamics alone, it becomes increasingly challenging to distinguish which one is more plausible than the others to represent the history of the Solar system. Therefore, some studies reversed their approach and began their analysis from the properties of old extraterrestrial solids that land on Earth - meteorites. Precise isotopic measurements of meteorites over the past decade provide further constraints on the meteorites' parent bodies' accretion history and timing of core formation (e.g. Dauphas, 2017; Dauphas and Pourmand, 2011; Kleine et al., 2002, 2009; Tang and Dauphas, 2014; Yin et al., 2002; Yu and Jacobsen, 2011), silicate differentiation (e.g. Borg et al., 2016; Kruijer et al., 2017b), magma ocean crystallization and crustal formation (e.g. Bouvier et al., 2018; Debaille et al., 2007; Harrison et al., 2008) and the parent bodies' bulk composition (e.g. Dauphas, 2017; Fitoussi et al., 2016; Toplis et al., 2013).

Among the terrestrial planets and their satellites, known samples are only available from Earth, the Moon and Mars, and most likely dwarf planet 4 Vesta. Hence, we can only constrain the accretion time and bulk composition of the two outer terrestrial planets, but not the inner two. Short lived radioactive decay systems, in particular the $^{182}$Hf-$^{182}$W system (half life = 8.9 Myr , Vockenhuber et al. (2004)), are widely adopted in estimating the core formation time and hence the accretion time scale of Earth and Mars (e.g.

Dauphas and Pourmand, 2011; Kleine et al., 2009; Kruijer et al., 2017b) because the daughter isotope is more siderophile than the parent isotope. In a review paper Kleine et al. (2009) found that Earth's accretion and core formation lasts longer than 30 Myr based on the Hf-W chronology (cf. Schiller et al., 2020, 2018). Mars, on the other hand, accreted much faster. Dauphas & Pourand (2011) showed that Mars could have accreted about half of its mass within $1.8^{+0.9}_{-1.0}$ Myr after the formation of calcium-aluminum rich inclusions (CAIs), which happened ca. 4.568 billion years ago (Bouvier & Wadhwa 2010). Such extremely rapid accretion was recapitulated by applying the $^{60}Ni$ - $^{60}Fe$ decay system (Tang and Dauphas, 2014). However, a recent study by Marchi et al. (2020) based on impact simulations suggest that Mars' accretion timescale could be as long as 15 Myr. Either way, the result is the same: this rapid timescale of the formation of Mars should be reproduced by the N-body simulations.

Isotopic measurements on both differentiated (achondrites) and non-differentiated meteorites (chondrites) provide insight for the initial composition of the initial planetesimals disc. Samples from the inner Solar system show a correlation between $^{54}Cr$ vs $^{50}Ti$ (Mezger et al., 2020 and references therein; Yamakawa et al., 2010) and $^{48}Ca$ vs $^{50}Ti$ (Dauphas et al., 2014; Warren, 2011). This begs the question whether a spatial isotopic gradient existed or could have been developed right before the formation of planetesimals (Yamakawa et al., 2010; cf. Schiller et al., 2018). Combining this suggested inner solar system isotopic gradient with N-body simulations, Mah and Brasser (2021) argue that Earth, Mars and Vesta have to have accreted from a narrow feeding zone close to their current locations in order to preserve the isotopic differences between these objects that we observe today. Their results thus favour planet formation models with less mixing throughout the disc (the classical and the depleted disc model) more than models involving giant planet migration through the asteroid belt (e.g. Grand Tack model) or inward drifting pebbles which enhance the disc's mixing.

### 1.5. Novel points and outline of this work

In this project, we want to re-investigate the formation of the terrestrial planets from the ground up, starting from a swarm of planetesimals in a manner similar to Kokubo and Ida (1998, 1995). We further want to go beyond just the dynamics and anchor the validity of our simulations and the chosen parameters to the cosmochemical chronology of the formation of Mars. We aim to simulate both the runaway phase and the oligarchic phase followed by the giant impact phase using the GPU acceleration N-body code *GENGA* (Grimm and Stadel, 2014). Compared to previous studies we do not apply artificial tricks which increase particles' radii to speed up the growth of planets (e.g. Carter et al., 2015; Kokubo and Ida, 1998), divide the initial disc into different annuli at the pre-runaway stage (Clement et al., 2020), or apply statistical methods to calculate the evolution of planetesimals during the pre-runaway stage (e.g. Walsh and Levison, 2019). Instead, we brute force the whole simulation starting from a disc of planetesimals before the pre-runaway stage by a direct N-body method and assume perfect accretion upon collision. We make use of a variety of Nvidia GPUs, with the highest-resolution simulations running on Tesla V100 GPUs at the University of Zurich. The Tesla V100 GPUs are ~4 to ~12 times faster than the Tesla K20 GPUs that have been previously used to simulate terrestrial planet formation (Clement et al. 2020).

In our first paper we present the first 10 Myr of our simulations of embryo formation, which is computationally the most challenging part. We study the classical and the depleted disc models, both of which do not involve migration of giant planets which limits the mixing of the disc - as suggested by the isotopic gradient of the inner disc (Mah and Brasser, 2021). In the next section, we briefly describe the N-body code *GENGA* and the effects of the gas disc we adopted in this study. We then present and discuss our results when assuming a gas disc with decay timescale of 2 Myr in Section 3 by mainly focusing on three aspects: (1) Mass distribution of the system; (2) embryo's formation timescale, and finally (3) accretion zones of the embryos. In Section 4, we showed that lowering the gas disc's decay timescale to 1 Myr, assuming embryo forming in a gas-free environment or assuming more circular orbits for the gas giants have decisive impacts on embryos formation timescales as well as accretion zones. We summarize our findings in the final section.

## 2. Method

### 2.1. Description of *GENGA* and the gas disc

We first briefly describe the N-body code *GENGA* and the effects from the gas disc. Developed based on another N-body code *MERCURY* (Chambers, 1999), Grimm and Stadel (2014) present a hybrid symplectic N-body code adopting GPU acceleration for simulating planetary systems with good energy conservation. It runs on all Nvidia's Graphics cards. *GENGA* supports three gravity modes for particles: 1) Fully interactive, all particles interact with every other particle. 2) Test particle mode, small particles only feel the gravity of large particles but don't perturb other particles. 3) Semi active mode, small particles interact with large particles, but not with other small particles. By using the gravity modes 2 or 3, the number of force term calculations can be reduced a lot, but also the simulation precision is reduced and small particle collisions are not handled correctly. In this work we always use the full interactive gravity mode. *GENGA* runs entirely on the GPU, in order to reduce the CPU-GPU data transfer bottleneck. On modern GPUs, *GENGA* can run simulations with several ten thousands of fully interactive particles. The big challenge of running high resolution N-body simulations as presented in this work, is the fact that at every time step a very large number of mutual close encounters has to be resolved, by still fulfilling energy conservation. *GENGA* contains several specialized and self-tuning GPU kernels to increase the performance as much as possible.

The gas disc implemented in *GENGA* followed from Morishima et al. (2010). Gas drag and Type I migration are included in the orbital evolution of particles. We briefly summarize the surface density profile of the gas disc and its effects on particles.

The gas surface density dissipated exponentially in space and time as:

$$\Sigma_{\mathrm{gas}}(r,t) \ = \ \Sigma_{\mathrm{gas,0}}(r\ /\ 1\ \mathrm{AU})^{-p}\ \exp\left(-t/\tau_{\mathrm{decay}}\right) \tag{1}$$

where where $\Sigma_{\mathrm{gas,0}}$ is $\Sigma_{\mathrm{gas}}$ at 1 AU and $t = 0$ and $\tau_{\mathrm{decay}}$ is the time scale for gas decay. $\Sigma_{\mathrm{gas,0}}$ is assumed to be 2000 $\mathrm{gcm^{-2}}$ and $p = 1$, which is the nominal disc from Morishima et al. (2010). This disc has a higher surface density than the MMSN over most of its range.

The aerodynamic drag force per unit mass (or acceleration) is given by (Adachi et al., 1976):

$$\boldsymbol{f}_{\rm drag} = (-1/2m)c_{\rm D}\pi r_{\rm p}{}^2\rho_{\rm gas}|v_{\rm rel}|v_{\rm rel} \tag{2}$$

where $m$ and $r_{\rm p}$ are the mass and the radius of a particle, $c_{\rm D}$ is the numerical coefficient which is assumed to be 2, $\rho_{\rm gas}$ is the gas density, and $\boldsymbol{v}_{\rm rel} = \boldsymbol{v} - \boldsymbol{v}_{\rm gas}$ is the velocity of a particle relative to gas velocity.

When the eccentricity and inclination of a planetesimal are both zero ($e = i = 0$), its semi-major axis damping time scale through TypeI-migration follows Tanaka et al. (2002) and is written as

$$\tau_{\rm tid1} = -\frac{a}{\dot{a}} = (2.7 + 1.1p)^{-1}\frac{M_\odot}{m}\frac{M_\odot}{\Sigma_{\rm gas}a^2}(\frac{c}{v_{\rm kep}})^2\Omega_{\rm kep}{}^{-1} \tag{3}$$

where $M_\odot$ is the Solar mass, $c$ is the isothermal sound velocity, $v_{\rm kep}$ is the Keplerian velocity of the particle and $\Omega_{\rm kep}$ is the Keplerian mean motion of the particle.

When $e$ and $i$ are non-zero, the mean evolution rate of the planetesimals' $e$ and $i$ is given by

$$\frac{\dot{e}}{e} = -\frac{0.780}{\tau_{\rm wave}} \tag{4}$$

$$\frac{\dot{i}}{i} = -\frac{0.544}{\tau_{\rm wave}} \tag{5}$$

where $\tau_{\rm wave}$ is the characteristic time of orbital evolution and is given by

$$\tau_{\rm wave} = \frac{M_\odot}{m}\frac{M_\odot}{\Sigma_{\rm gas}a^2}(\frac{c}{v_{\rm kep}})^4\Omega_{\rm kep}{}^{-1} \tag{6}$$

which is shorter than the orbital decay time $\tau_{\rm tid1}$ obtained in Tanaka et a. (2002) for $e = i = 0$ (equation 3) by a factor of $(c/v_{\rm kep})^2$.

Equation 4 and equation 5 are only limited to small $e$ and $i$ ($\ll H/a$, where $H = c/\Omega_{\rm kep}$ is the scale height) (Tanaka and Ward, 2004), or that the particles' motion relative to gas remain subsonic. Papaloizou and Larwood (2000) found that the time scales in equation 3 and equation 6 increase with increasing $e$ by factors $g_{\rm tid1}$ and $g_{\rm wave}$, respectively, from a two dimensional linear theory:

$$g_{\rm tid1} = \frac{1+(x/1.3)^5}{1-(x/1.1)^4}, \tag{7}$$

$$g_{\rm wave} = 1 + \frac{1}{4}x^3, \tag{8}$$

The original $x$ used in Papaloizou and Larwood (2000) is $x_{\rm PL} = ae/h$. Morishima et al. (2010) modify this for non-zero $i$ in order to reproduce similar results with the hydrodynamic simulations conducted by Cresswell et al. (2007). The modified $x$ is hence

$$x = \frac{a\sqrt{(e^2+i^2)}}{2h}. \tag{9}$$

The modified eccentricity and inclination evolution for large $e$ and $i$ are then written as:

$$\frac{\dot{e_{\rm tid}}}{e} = -\frac{0.780}{\tau_{\rm wave} g_{\rm wave}} \tag{10}$$

$$\frac{\dot{\iota_{\rm tid}}}{i} = -\frac{0.544}{\tau_{\rm wave} g_{\rm wave}} \tag{11}$$

These equations correspond to equation 4 and equation 5, but are multiplied by $g_{wave}^{-1}$ from equation 8. From equation 3, equation 7 and equation 10, the evolution of $a$ in large $e$ case is given by

$$\frac{\dot{a}_{\rm tid}}{a} = -\frac{1}{\tau_{\rm tid1} g_{\rm tid1}} - \frac{\dot{e_{\rm tid}}}{e}\left(\frac{2e^2}{\sqrt{1-e^2}}\right) \tag{12}$$

where the first term in the right hand side comes from exchange of angular momenta while the second term represents the migration rate due to energy dissipation .

*GENGA* allows to choose if the gas drag and the gas disc is applied to all particles or to exclude the gas giants from its influence. In this work, we exclude the gas giants from the gas drag and gas disc potential.

## 2.2. Initial conditions

We run simulations for both the classical model (e.g. Chambers, 2001; O'Brien et al., 2006; Raymond et al., 2006, 2009) and the depleted disc model (Izidoro et al., 2014; Mah and Brasser, 2021). In the classical model, the solid surface density profile $\Sigma_{\rm s}$ = 7 gcm$^{-2}$ ($a$/ 1 AU)$^{-3/2}$, which follows the MMSN (Hayashi, 1981). In the depleted disc model, the solid surface density profile is given as the following:

$$\Sigma(a) = \begin{cases} \chi\,\Sigma_{\rm s}\,(a/\,1\,{\rm AU})^{-3/2} & a \leq a_{\rm dep} \\ (1-\beta)\,\chi\,\Sigma_{\rm s}(a/\,1\,{\rm AU})^{-3/2} & a > a_{\rm dep} \end{cases} \tag{13}$$

where $a_{\rm dep}$ = 1, 1.25 or 1.5 AU is the depletion location, $\beta$ = 50% or 75% is the depletion factor and $\chi$ = 1, 1.5 or 2 is the scaling factor of MMSN. The parameters are chosen from simulations in Mah and Brasser (2021) that achieved the highest probability of forming Mars with the correct mass. The local depletion at the further region of the protosolar disc has been suggested to be generated due to difference in disc viscosity in inner and outer region (Jin et al., 2008) leading to a different drift rate of dust between the two regions and hence redistribution of the solid surface density (e.g. Drążkowska et al., 2016). Figure 1 shows the solid surface density profile of the depleted disc model adopted in our study compared to the MMSN for the classical model. In all the cases, the depleted disc model has a lower initial solid surface density than the MMSN in the current Mars and the asteroid belt region ($a \geq 1.5$ AU). Table 1 shows the simulation sets we performed for both models. We distribute equal-mass planetesimals at the beginning of the simulation in between 0.5 to 3 AU according to the solid surface density profile described above and in table 1. In most of our runs, the initial radius of a planetesimals is set to be 350 km, but we also performed many lower resolution simulations with $r$ = 800 km for comparison (LR). The density of planetesimals is set to be 3 g cm$^{-3}$. All the planetesimals are fully self-gravitating. The high-resolution simulations typically began with ~30,000 to 40,000 planetesimals while the low-resolution simulations began with ~2000 to 3000. The eccentricities and inclinations of the planetesimals are uniformly random in the range $0 < e < 0.01$ and $0^{\rm o} < i < 0.5^{\rm o}$. Their nodal angles and mean anomalies are uniformly random from 0$^{\rm o}$ to 360$^{\rm o}$. We include the gas disc profile and its effects following Morishima et al. (2010) as described in the previous section, except for two sets of control simulations. The timescale of the decline of the gas disc, $\tau_{decay}$, is either set to be 1 Myr (T1) or 2 Myr (default). In all of our simulations the gas giants do not feel the potential of the gas disc.

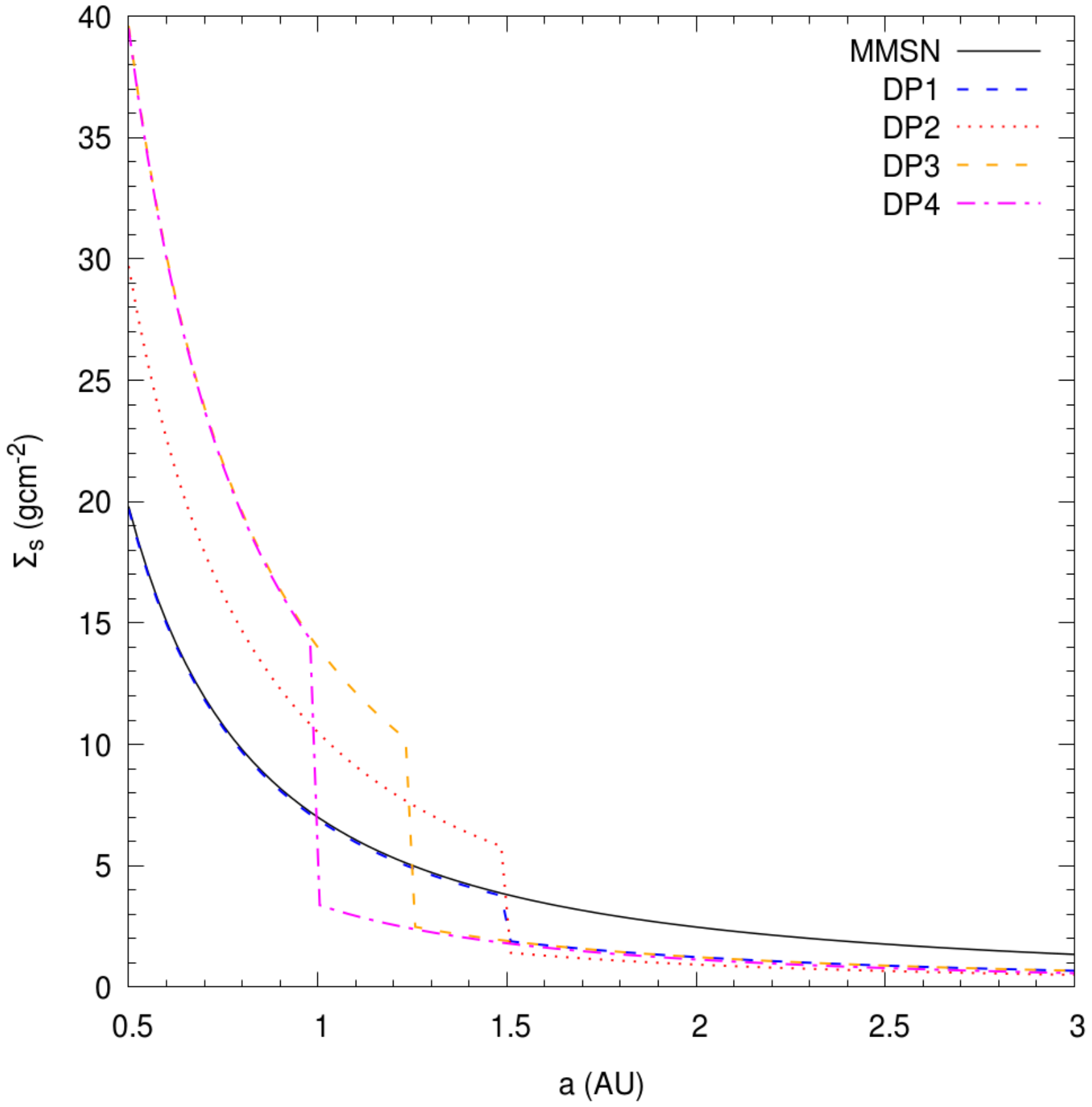

Figure 1 - The solid surface density profile of the depleted disc model, compared to the minimum mass Solar nebula (MMSN) adopted for the classical model. DP1, DP2, DP3 and DP4 refers to the different initial conditions of the depleted disc model (see table 1.).

Table 1 - The initial conditions of our classical and depleted disc simulations, as well as simulations of Mars region annulus. We study the growth of embryos from a disc of planetesimals with different simulation set-ups, including planetesimals' radii ($r$) , number of initial planetesimals ($N$), total mass of the initial disc, decay timescale of the gas disc ($\tau_{decay}$) and orbits of the giant planets - eccentric Jupiter-Saturn (EJS) and circular Jupiter-Saturn (CJS). The GPU time of the other cards is calculated according to the following relation: (GPU time in V100) = (GPU time in P100)/2 = (GPU time in K20)/4.

| | Name of simulation sets | $a_{dep}$, $\beta$, $\chi$ (Equation 13 ) | Planetesimals' radii ($r$) | No. of initial planetesimals ($N$) | Mass of initial disc (in $M_{Earth}$) | $\tau_{decay}$ (Equation 1) | Giant planets' orbit | Number of simulations | Time taken per simulation in V100 card (hours) |
|---|---|---|---|---|---|---|---|---|---|
| Depleted Disc | DP1 | 1.5 AU, 50%, 1 | 350 km | 28168 | 2.5 | 2 Myr | EJS | 2 | 885.7 |
| | DP2 | 1.5 AU, 75%, 1.5 | 350 km | 35307 | 3.2 | 2 Myr | EJS | 2 | 859.4 |
| | DP3 | 1.25 AU, 75%, 2 | 350 km | 41235 | 3.7 | 2 Myr | EJS | 2 | 914.7 |
| | DP4 | 1 AU, 75%, 2 | 350 km | 34763 | 3.1 | 2 Myr | EJS | 2 | 866.4 |
| | DP1_LR | 1.5 AU, 50%, 1 | 800 km | 2337 | 2.5 | 2 Myr | EJS | 5 | 251.6 |
| Classical | CL1 | NIL | 350 km | 37437 | 3.4 | 2 Myr | EJS | 2 | 1399.8 |
| | CL2 | | 350 km | 37437 | 3.4 | 2 Myr | CJS | 2 | 1373.3 |
| | CL1_nogas | | 350 km | 37437 | 3.4 | NIL | EJS | 2 | 1469.2 |
| | CL1_LR | | 800 km | 3114 | 3.4 | 2 Myr | EJS | 5 | 252.0 |
| | CL2_LR | | 800 km | 3114 | 3.4 | 2 Myr | CJS | 5 | 186.6 |
| | CL1_nogas_LR | | 800 km | 3114 | 3.4 | NIL | EJS | 3 | 271.5 |
| | CL1_T1 | | 350 km | 37437 | 3.4 | 1 Myr | EJS | 2 | 1298.7 |
| | CL1_LR_T1 | | 800 km | 3114 | 3.4 | 1 Myr | EJS | 5 | 206.0 |
| Mars region annulus | Annu_125km | NIL | 125 km | 32755 | 0.12 | 2 Myr | EJS | 2 | 1421.5 |
| | Annu_500km | | 500 km | 508 | 0.12 | 2 Myr | EJS | 2 | 99.9 |
| | | | | | | | | Total time | 27976.3 |

The initial orbits of the giant planets are set to be either having their eccentric and inclined current orbits (EJS: $a_j$ = 5.20 AU, $e_j$ = 0.049, $i_j$ = 0.33$^o$ ; $a_s$ = 9.58 AU, $e_s$ = 0.056, $i_s$ = 0.93$^o$, where ($a_j$,$e_j$, $i_j$) and ($a_s$,$e_s$,$i_s$) are the semi-major axis, eccentricity and inclination of Jupiter and Saturn, respectively), or more circular and planar orbits adopted in the Nice model (Tsiganis et al., 2005) (CJS: $a_j$ = 5.45 AU, $e_j$ = 0.0009, $i_j$ = 0$^o$ ; $a_s$ = 8.18 AU, $e_s$ = 0.0002, $i_s$ = 0.5$^o$). Compared to Walsh and Levison (2019) and Clement et al. (2020), we introduce the giant planets in a more early stage since we would like to study how the gravitational influence of the gas giants affect the formation of embryos. A rapid formation of Jupiter and Saturn could in principle be explained by fragmentation that forms gas clump in the gas disc, following by gravitational instability of the fragment within the protosolar disc (Boss, 1997; Durisen et al., 2007; Vorobyov and Elbakyan, 2018), instead of the conventional core-accretion (e.g. Pollack et al., 1996; Lissauer et al., 2009).

It is currently still unclear whether the giant planets possessed eccentric or circular orbits during the gas disc phase. Although the traditional "Nice model" assumes circular orbits (CJS) for the giant planets, some hydrodynamical simulations result in eccentric Jupiter and Saturn if they migrated in the gas disc and became trapped in mean-motion resonances (Pierens et al. 2014). Using the final orbital configuration of the gas giants from their hydrodynamical simulations, Pierens et al. (2014) claim that they successfully reproduce the outer solar system architecture. Also, if the giant planets' cores formed via pebble accretion, involving mutual scattering between planetesimals (Levison et al. 2015a), these cores and their fully formed giant planets may not be in very circular orbits ($e$ < 0.01). Furthermore, if gas giants formed from gravitational instability, their initial eccentricities are not well determined. Most previous high resolution simulations (Walsh & Levsion 2019; Clement et al. 2020) studied the CJS scenario, in contrast we place greater emphasis on the results from the EJS scenario while using CJS simulations for comparison and for statistical completeness**.**

We first present the results of our simulation with *GENGA* for the first 10 Myr, in which we mainly focus on the formation of embryos. The time step is set to 5 days, which is more than 25 steps for the innermost planetesimals. Particles are removed when they are closer to 0.1 AU from the Sun or further than 100 AU. All our high resolution simulations ($r$ = 350 km) are performed on Nvidia Tesla V100 cards on the Service and Support for Science IT (S$^3$IT) cluster of the University of Zurich, and Nvidia Tesla P100 cards on the Swiss National Supercomputing Centre (CSCS) Piz Daint supercomputer in Lugano, Switzerland. The low resolution simulations are performed mostly on Tesla K20 cards at the Earth Life Science Institute in Tokyo, Japan and they take about a month to finish. Table 1 also shows the computing time on the Tesla V100 cards, which are the fastest among the GPUs we had access to. We divide a factor of 2 from the total simulation time in P100 (i.e. V100 is ~2 time faster than P100) and a factor of 4 from the total simulation time in K20 (V100 is ~4 times faster than K20). The total GPU time in Tesla V100 cards for all simulations is 27976.3 hours.

## 3. Results with $\tau_{decay}$ = 2 Myr

### 3.1. Mass distribution

We first define objects into 3 categories - *embryos* with masses larger than 1 $M_{\mathrm{Moon}}$ ($M_{\mathrm{Moon}}$ is the lunar mass, which is ~ 0.012 $M_{\mathrm{Earth}}$, where $M_{\mathrm{Earth}}$ is the Earth's mass), *proto-embryos* with masses < 1 $M_{\mathrm{Moon}}$ but > $10^{-3}$ $M_{\mathrm{Earth}}$ (~11 times the initial masses of planetesimals and more than 5 times of Ceres' mass) and *planetesimals* with masses < $10^{-3}$ $M_{\mathrm{Earth}}$. We adopt different conditions from Clement et al. (2020) for the definition of proto-embryos and planetesimals because the initial masses of our planetesimals are larger than those in Clement et al. (2020). The three categories of objects are defined only for data analysis purposes, **so** that we can verify whether different regions of the disc achieve a bimodal mass distribution or not. In the N-body integration, there is no difference between them.

We also define the Mercury, Venus, Earth and Mars region as 0.27 AU < $a$ < 0.5 AU, 0.5 AU < a < 0.85 AU, 0.85 AU < $a$ < 1.25 AU and 1.25 AU < $a$ < 1.75 AU. These regions are defined for data analysis purposes. In the following we investigate three different aspects of the mass distribution. First, we focus on embryo growth as a function of semi-major axis and its dependence on the initial model (section 3.1.1). Second, we demonstrate the effects of Type-I migration (section 3.1.2), and finally we present the evolution of the bimodality between embryos and planetesimals (section 3.1.3).

#### *3.1.1. Embryo growth*

After 10 Myr, embryos have formed throughout the disc. Figure 2 shows the snapshot of the high-resolution classical (CL1) and the depleted disc models (DP1) in terms of mass versus semi-major axis of the objects in the disc; Jupiter and Saturn were on their current eccentric orbits (EJS) embedded in a gas disc, with an e-folding time of $\tau_{decay}$ = 2 Myr (equation 1). In the first 0.01 Myr, planetesimals remain small and do not grow beyond 4 times their initial mass. In the next 0.1 Myr, collisions occur frequently and sub-lunar sized proto-embryos (blue circles) emerge at the inner edge of the disc. Still nothing grows above a lunar mass at this stage. After a million years of collisional growth, embryos (black circles) with masses > 1 $M_{\rm Moon}$ appear within 1 AU in both models. Beyond 1.5 AU, however, everything still remains as small planetesimals (red circles). Once the embryos have grown to 2 - 3 $M_{\rm Moon}$, they start to migrate inward due to Type-I migration. This results in a convoy of embryos within 0.5 AU, which is closer than the original inner edge of our solid disc. Several proto-embryos (blue circles) are also located within 0.5 AU due to scattering by embryos instead of Type-I migration.

In principle, embryos should stop growing by accreting small planetesimals after reaching the isolation mass, which is a theoretical upper limit of mass for embryos that fully accreted all the mass within their feeding zone. The isolation mass, which is represented by the dotted line in figure 2 is written as

$$M_{\rm iso} = 2\pi ab \sum(a) \tag{14}$$

where $a$ is the semi-major axis of the embryo, $b$ is the width of the embryo's feeding zone, which is typically 10 mutual Hill radii of the two adjacent embryos (Kokubo and Ida, 1998) and $\Sigma$ is the solid surface density within its feeding zone. We find that objects within 1 AU begin migrating inward due to Type-I migration before reaching the isolation mass, especially for those embryos undergoing runaway growth at 1 AU. While migrating they reach $M_{\rm iso}$ in the current region of Mercury and Venus because of the dependence of $M_{\rm iso}$ on distance. Embryos that frist formed within the Mercury and Venus region mostly do not survive until the end of simulation due to the long-lived gas disc, and they are lost to the Sun. Hence, the embryos within 0.5 AU we see after 5 Myr are not the same embryos within 0.5 AU after 1 Myr since the latter are lost to the Sun.

In the Mars region embryos do not emerge until 5 Myr, with 5 of them located beyond 1.5 AU in the classical model. However, no embryos formed in the depleted disc at this time. This suggests that we may have a problem of forming Mars quick enough at 1.5 AU in the depleted disc model to match its Hf-W and Fe-Ni chronological data (Dauphas and Pourmand, 2011; Tang and Dauphas, 2014), despite the fact the depleted disc model has a higher chance of forming a "small" Mars (Izidoro et al., 2014; Mah and Brasser, 2021) (See Section 3.3 for further discussion).

Up to this point (5 Myr), the asteroid belt region is still in a quiescent state with very little growth. Proto-embryos with more than 10 times the mass of the current largest asteroid, Ceres, could form in the classical model but not in the depleted disc model. This is due to the lower initial solid surface density assumed in the depleted disc. Before they can grow further, the dissipation of the gas disc leads to an increase in the precession rate of the argument of perihelion planetesimals. When the precession rate of the planetesimals (or proto-embryos) matches that of Jupiter, a secular resonance occurs which increases the eccentricity of the planetesimals (Murray and Dermott, 1999). Since the gas disc continues to dissipate and the precession rate of the planetesimals keeps changing accordingly, the location of the secular resonance with Jupiter moves inward to the Sun with time. This leads to the "sweeping secular resonance" effect (Bromley & Kenyon 2017; Heppenheimer, 1980; Hoffmann et al., 2017; Lecar and Franklin, 1997; Nagasawa et al., 2000, 2001; Ward, 1981) that clear the planetesimals in the asteroid belt region (Morishima et al., 2010; Nagasawa et al., 2005; Zheng et al., 2017) without the requirement of the migration of the giant planets (e.g. Grand Tack model, Walsh et al. (2011). In our simulation wherein $\tau_{decay}$ = 2 Myr this effect starts taking place at 5 Myr and continues until the end of the simulation. Therefore, at the end of the simulation (10 Myr) we have a nearly empty asteroid belt. On the other hand, the inner region ($a$ < 1.8 AU) has 20 to 30 embryos, with Mars-sized embryos at 1.5 AU only existing in the classical model.

Our simulations reproduce a series of lunar- to Mars-sized embryos from the region spanning the current orbits of Mercury to Mars as also observed in Walsh & Levison (2019). However, since Walsh & Levison (2019) assumed more circular orbits for Jupiter and Saturn (CJS), their simulations do not show visible effects of a sweeping secular resonance and hence in their simulations embryos also exist beyond Mars' orbit at 10 Myr. In contrast, our simulations (figure 2) adopt the EJS scenario, resulting in a much stronger mass depletion effect due to the sweeping secular resonance. Thus we do not have any embryos existing in the region beyond Mars at 10 Myr. Embryos which originated in the region beyond Mars are mostly transferred to the inner disc via the sweeping secular resonance.

Our simulation also produces a population of 5 to 7 embryos that are stranded within 0.5 AU in both models at 10 Myr, but with almost no planetesimal in that region (figure 2). This is one of the first results indicating that the Mercury-region achieves a high embryos' to planetesimals' mass ratio of > 50 right after the embryo formation stage.

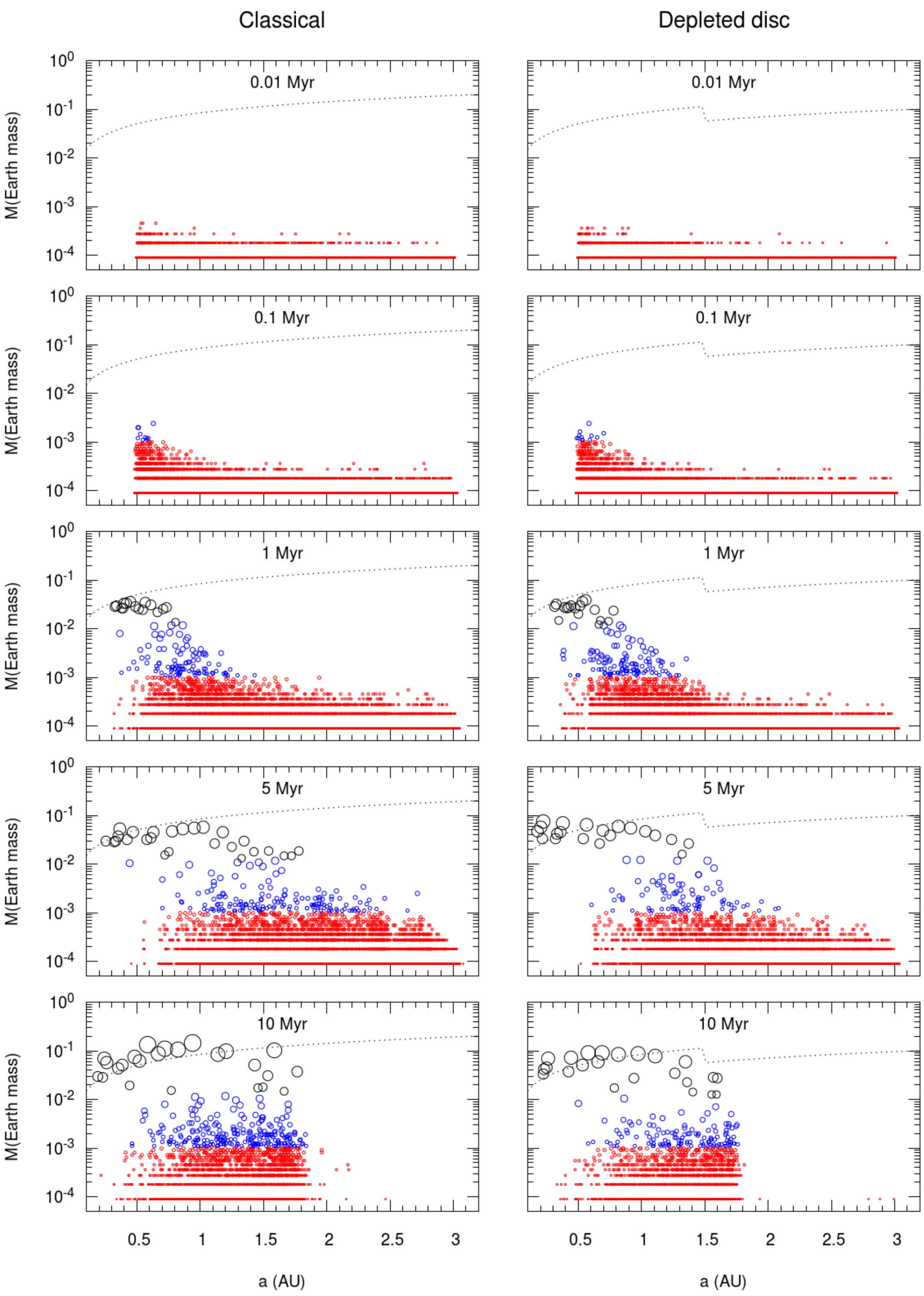

Figure 2 - Mass versus semi-major axis of embryos (black circle), proto-embryos (blue circle) and planetesimals (small red dots) at 0.01, 0.1, 1, 5, and 10 Myr. The left panels show results of the classical model (CL1) with initially eccentric Jupiter and Saturn (EJS). The right panels show results of the depleted disc model (DP1) (see table 1 for the initial conditions). The initial radii of the planetesimals are 350 km and the gas disc's $\tau_{decay}$ = 2 Myr for both models. The dotted line represents the isolation mass of the embryo.

#### *3.1.2. Effect of Type-I migration*

The long-lived gas disc has a decisive impact on the embryos' growth and evolution. We observe that at 10 Myr some embryos in the Mercury-Venus region indeed originated from around the Earth region. Figure 3 shows semi-major axis (upper panel) and mass growth (lower panel) of the most massive embryos in the Mercury, Venus, Earth and Mars regions at 10 Myr of the classical high resolution EJS simulation (CL1). The first generation of embryos form in the innermost region (0.5 to 0.7 AU) within 1 Myr of the simulation by the time the gas disc is still thick. The timescale of Type I migration is shorter than the time scale of the embryos' growth and hence these first generation of embryos were lost to the Sun. The Mercury and Venus embryos found in this simulation are the next generation embryos which form slower than the first generation; most of these form in the Earth region during the first 2 Myr. They then migrate from 1 AU to the Mercury and Venus region in the next 5 Myr. Their migration ceased at around 7 Myr when the gas disc's surface density has further dissipated after more than 3 e-folding times. Unlike the embryos in the Mercury and Venus region, figure 3 shows that the most massive embryos in the Earth and Mars region did not migrate to the innermost region. The embryos in the Earth region form slightly slower than those that migrate to the Mercury and Venus regions, and reach the Type-I migration mass (~2 to 3 $M_{\mathrm{moon}}$) later. Since these embryos reach the Type-I migration mass after about 5 Myr, by that time the gas has already dissipated by more than 2 e-folding times, so that the embryos only migrate inward by < 0.1 AU . A similar situation also applies to the Mars region. With an even slower growth, the embryos in the Mars region do not undergo significant Type-I migration throughout the 10 Myr simulation. Instead, they are scattered from 1 AU to 1.5 AU and reach 1 martian mass ($M_{\mathrm{Mars}}$) only after 10 Myr. In this case, we might expect Mars to form from similar material that forms Earth. However, the nucleosynthetic isotopic data from martian meteorites (e.g. Lodders and Fegley, 1997; Sanloup et al., 1999; Tang and Dauphas, 2014; Brasser et al., 2017) argue against this martian formation route (Brasser et al., 2018; Woo et al., 2018).

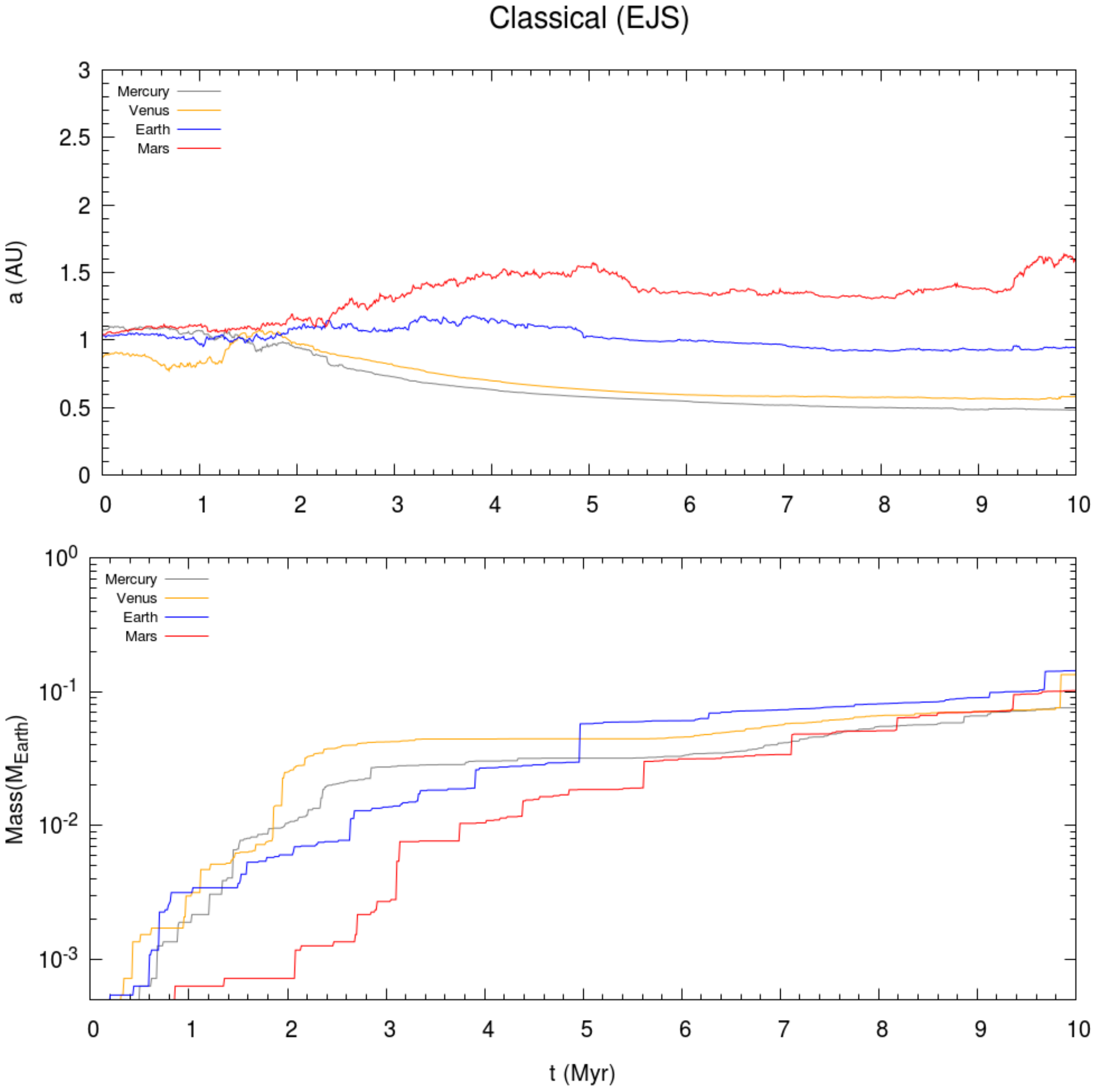

Figure 3 - Time evolution of the semi-major axes (upper panel) and masses (lower panel) of the most massive Mercury, Venus, Earth and Mars region embryo at 10 Myr in the classical model (CL1) with initially eccentric Jupiter and Saturn (EJS). The initial radii of planetesimals are 350 km and the gas disc's $\tau_{decay}$ = 2 Myr.

#### *3.1.3. Mass evolution of the system*

As the embryos grow, the mass distribution of the system changes with time. Previous studies of terrestrial planets' formation usually begin with a bimodal mass distribution (e.g. Brasser et al., 2016; Clement et al., 2018, 2019; Jacobson and Morbidelli, 2014; O'Brien et al., 2006; Raymond et al., 2006, 2009; Walsh et al., 2011) of the system throughout the whole disc at a particular time after the oligarchic growth phase (Kokubo and Ida, 1998, 2000, 2002; Chambers, 2006). Walsh and Levison (2019) have shown that the oligarchic growth phase is not achieved at the same time in different regions of the disc, and the mass distribution of the disc is continuous in the asteroid region after 20 Myr (i.e. non-bimodal). Here, we further support their results by showing the time evolution of the mass distribution of the system with time for our high resolution simulations in figure 4. We found that the classical and the depleted disc simulations' results are similar. The percentage of mass in embryos quickly increases to ~80 % within 2

Myr, indicating the fast growth of embryos in the inner region due to the shorter orbital period and higher surface density of planetesimals. Intriguingly, the mass percentage in embryos stays at about 80%, which indicates that the inner region has never been fully depleted in small planetesimals and proto-embryos. They are continuously supplied to the inner region from the outer regions through gas drag driven migration, mutual scattering between planetesimals and embryos and later by the sweeping secular resonance (See Section 3.2 for further description). The loss of embryos to the Sun through Type-I migration is the other reason why the mass percentage in embryos never reaches over 90% within 10 Myr. On the other hand, the mass distribution in the region 1 AU < $a$ < 2 AU is very different from that where $a$ < 1 AU. Embryos, although continuously increasing in mass percentage, consist of less than 40 % of mass within 10 Myr, which means that many objects remain sub-lunar in mass. Intermediate-mass objects (proto-embryos) make up at least 20 % of the total mass after 5 Myr, suggesting that the mass distribution within this region is non-bimodal within the first 10 Myr. The region beyond 2 AU has more than 80% of the total mass in small planetesimals at all times. Hence our results reproduce the strong inside-out growth seen in previous studies using different initial conditions (Morishima et al. 2010, Carter et al. 2015, Clement et al. 2020 and Walsh & Levison 2019). Our results do not show a bimodal mass distribution of embryos and planetesimals throughout the whole disc. In contrast to the claim of Walsh & Levision (2019), we do not attribute the lack of embryo formation in the asteroid belt to collisional grinding, but only to the longer orbital period, the lower solid surface density and, in the EJS scenario, to the sweeping secular resonance.

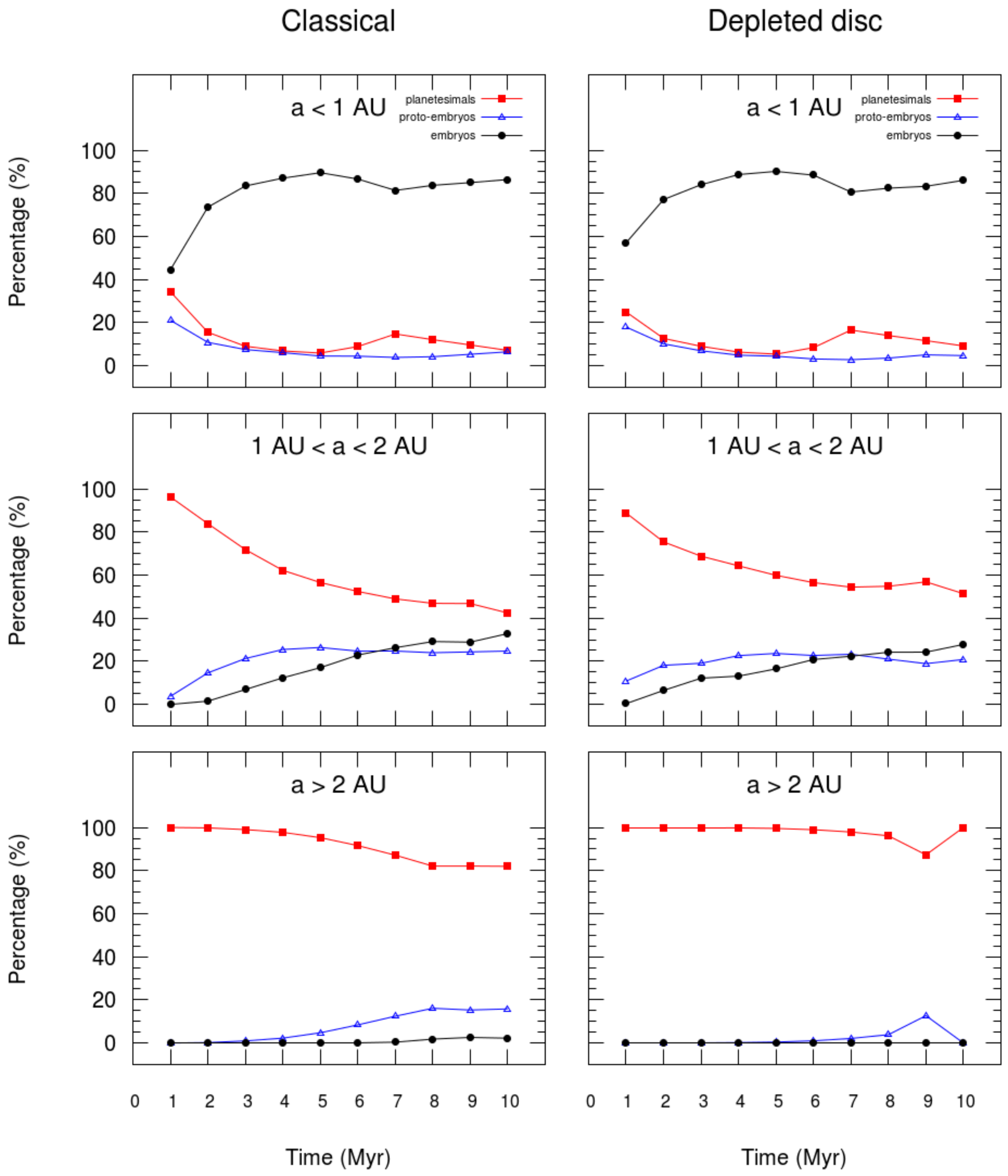

Figure 4 - Change of the percentage of mass in embryos (black circle), proto-embryos (blue triangle) and planetesimals (red square) with time in the region of $a < 1$ AU, 1 AU $< a <$ 2 AU and $a >$ 2 AU, where a is the semi-major axis. The left panels show statistics of all the classical model simulations with initial planetesimals radii = 350 km (CL1 + CL2) within a gas disc with $\tau_{decay}$ = 2 Myr. The right panels are the same as the left panels, but only for the depleted disc model (DP1 - DP4). Embryos are defined as objects having masses > lunar mass (~0.012 $M_{Earth}$), proto embryos are objects with masses in range of $10^{-3}$ $M_{Earth}$

to lunar mass (~ 11 times of the initial mass of planetesimals and 5 times larger than Ceres' mass) and planetesimals are defined as object less than $10^{-3}$ $M_{Earth}$.

### 3.2. Composition of embryos

In the previous section, we showed that the sweeping secular resonance depleted the asteroid belt region without requiring the giant planets to migrate through the asteroid belt. Intriguingly, the materials in the asteroid belt that are swept up do not get ejected from the system, but are implemented into the inner solar system region. Although the effect of the sweeping secular resonance has been recognized in previous N-body simulation results (e.g. Bromley & Kenyon 2017**;** Hoffmann et al., 2017; Morishima et al., 2010; Zheng et al., 2017), their effect on the embryos' or planets' final feeding zones and bulk composition has yet to be examined. In this section, we briefly review how the sweeping secular resonance occurs and show that how it alters the composition of embryos.

#### *3.2.1 Effect of the sweeping secular resonance*

Once the precession rate of the argument of perihelion of a planetesimal matches the secular eigenfrequencies of one of the giant planets, the eccentricity of the planetesimal is rapidly increased. Figure 5 shows the snapshot of the high resolution EJS simulation (CL1) in a gas disc with $\tau_{decay}$ = 2 Myr. The orbits of the materials in the disc still remain dynamically cold at 3 Myr because of their eccentricities and inclination being damped by the gas disc. As the gas disc's surface density is continuously decreasing, it increases the precession rate of the argument of perihelion of the planetesimals ($d\varpi/dt$). The outer region of the disc first matches the secular eigenfrequency of Jupiter ($g_5$) and this leads to a secular resonance ($\nu_5 = d\varpi/dt - g_5 = 0$) which pumps up the eccentricities of the planetesimals at around 2.8 AU as we observe at 6 Myr. We computed the analytical location of $\nu_5$ from the appendix of Nagasawa et al. (2000) and is shown as the vertical dashed line in figure 5. We do not plot the secular resonance of Saturn ($\nu_6 = d\varpi/dt - g_6 = 0$) because its location is always further away from the Sun than $\nu_5$ and hence its is relatively less effective in pumping the eccentricities of the planetesimals up than $\nu_5$. Once the eccentricities of the planetesimals are pumped up to large values (> 0.5), their orbits go through the inner region of the disc which has a much higher gas surface density. This damps down the high eccentricities of the planetesimals initially at approximately constant semi-major axis, followed by a phase of exchange between eccentricity and semi-major axis by conservation of angular momentum. Ultimately this loss of angular momentum brings the material to the inner solar system. We find that materials affected by the secular resonance are usually brought to the Mercury-Venus region in our simulations. As time goes on, the gas disc continues to dissipate and the $\nu_5$ resonance moves inward to around 2.3 AU at 8 Myr. This process continues to the end of our simulation (10 Myr) and most of the material beyond Mars' orbit are kicked up and brought to the Mercury-Venus region. Figure 6 shows the graphical illustration of the $\nu_5$ secular resonance effect at 8 Myr on the planetesimals' eccentricities and semi-major axes. Planetesimals eccentricities first gets pumped up to around 0.6 to 0.8 (green arrows pointing upward) by the $\nu_5$ resonance that sweeps sunward (green arrows pointing leftward). The eccentricities and semi-major axes of the planetesimals are then damped down due to gas drag (blue arrow pointing downward), leading to the inward implantation of planetesimals beyond 2 AU to the Mercury and Venus region (large transparent pink arrow). In this simulation the average implantation flux of the planetesimals is ~0.08 $M_{\text{Earth}}$ Myr$^{-1}$.

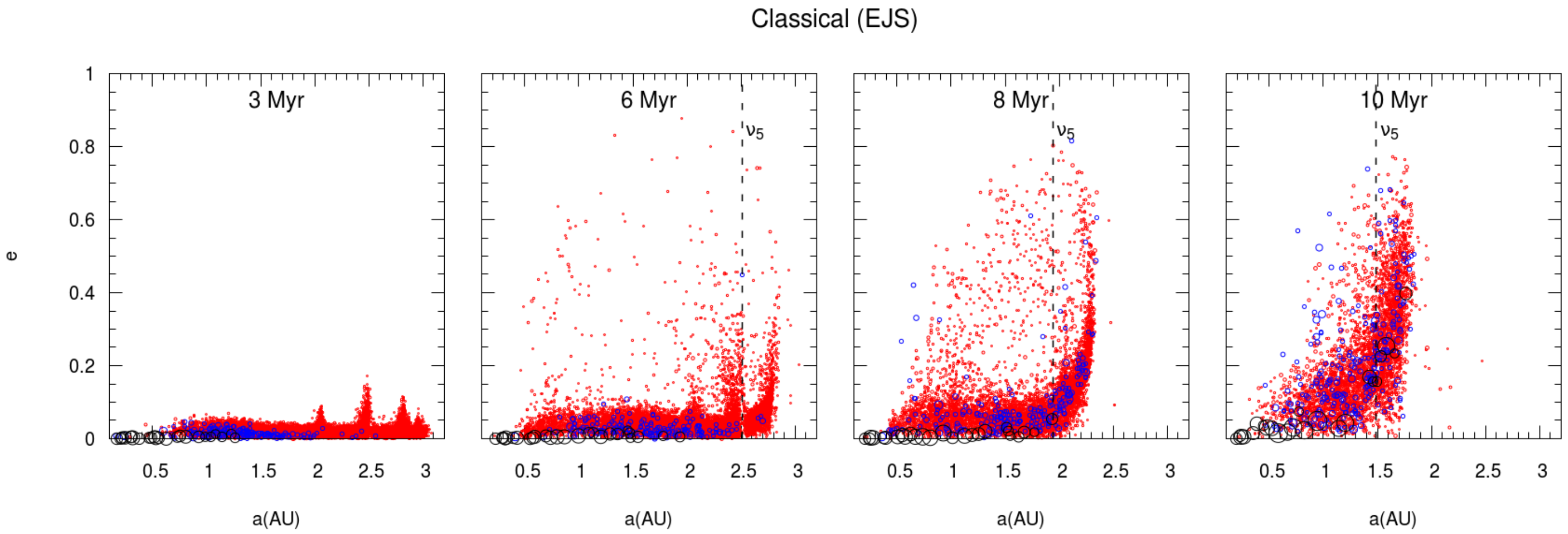

Figure 5 - Eccentricities versus semi-major axes of embryos (black large circle), proto-embryos (blue medium circle) and planetesimals (red small circle) at 3, 6, 8 and 10 Myr for the classical model (CL1) with initial eccentric Jupiter and Saturn (EJS). The initial radii of planetesimals are 350 km and the gas disc's $\tau_{decay}$ = 2 Myr. The vertical dotted line indicates the location of $\nu_5$ secular resonance computed from the appendix of Nagasawa et al. (2000).

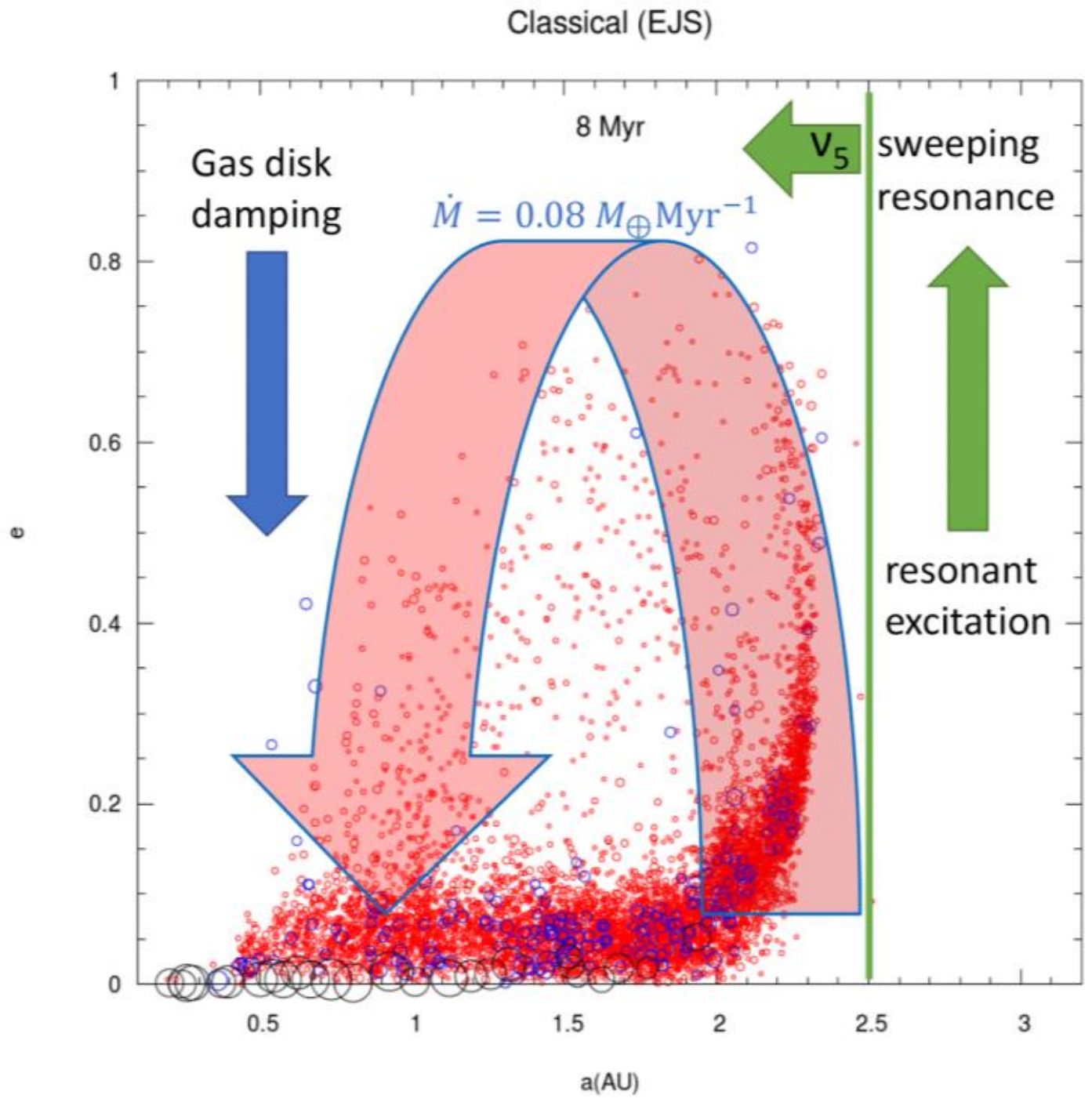

Figure 6 - Graphical illustration of the $\nu_5$ secular resonance effects at 8 Myr on the planetesimals' eccentricities and semi-major axes. This simulation is taken from the classical model (CL1) with initially eccentric Jupiter and Saturn (EJS) (figure 5). When $\nu_5$ resonance sweeps inward (green arrow pointing left), the eccentricities of planetesimals are excited to ~0.6 - 0.8 (green arrow pointing upward) at the resonance

location. The eccentricities and semi-major axes of the planetesimals are then damped down due to gas drag (blue arrow pointing downward), leading to the inward implantation of planetesimals originating from beyond 2 AU to the Mercury and Venus region (large transparent pink arrow). In this simulation the average implantation flux of the planetesimals is ~0.08 $M_{\text{Earth}}$ Myr$^{-1}$.

### *3.2.2 Accretion zones of embryos*

The secular resonance completely alters the accretion zone (or feeding zones) and hence the composition of the embryos formed in the terrestrial planet region. Figure 7 shows the accretion zones of embryos at 10 Myr that end up in the Mercury (grey), Venus (orange), Earth (blue) and Mars regions of the classical and the depleted disc simulations. We define the mass weighted mean initial semi-major axis, $a_{\text{weight}}$ as the mean of the accretion zone of each embryo. It is written as:

$$a_{\text{weight}} = \frac{\sum_i^{N_{\text{acc}}} m_i a_i}{\sum_i^{N_{\text{acc}}} m_i} \tag{15}$$

(Brasser et al., 2017; Fischer et al., 2018; Kaib and Cowan, 2015; Woo et al., 2018), where $m_i$ and $a_i$ are the initial mass and semi-major axis of the accreted planetesimals and $N_{acc}$ is the number of planetesimals accreted by the embryo. Its mass weighted standard deviation is written as

$$\sigma_{\text{weight}} = \sqrt{\frac{\sum_i^{N_{\text{acc}}} m_i (a_i - a_{\text{weight}})^2}{((N_{\text{acc}}-1)/N_{\text{acc}}) \sum_i^{N_{\text{acc}}} m_i}} \tag{16}$$

(Brasser et al., 2017; Kaib and Cowan, 2015; Mah and Brasser, 2021; Woo et al., 2018).

The accretion zone for each embryo in figure 7 is represented by $a_{\text{weight}} \pm \sigma_{\text{weight}}$. We combined the results from low resolution and high resolution simulations because their results are similar. We first noted that in both models, the accretion zones of all embryos are highly overlapping and most embryos' $a_{\text{weight}}$ are above the linear correlation line $y = x$. Both can be explained by the strong mixing effect during embryo formation due to the long lived gas disc causing long lasting inward migration of the embryos, and the sweeping secular resonance. In both models, many embryos in Mercury and Venus region have accretion zones extending beyond 2 AU, some of which even have $a_{\text{weight}} > 2$ AU. This is unexpected since embryos in the innermost solar system are expected to accrete material mainly in the inner region. As shown in figure 2 and figure 5, the sweeping secular resonance occurred beyond the Mars region and brought material beyond 2 AU to the Mercury-Venus region. This explains why they accreted more outer solar system material than expected and could be the reason for Venus' higher atmospheric nitrogen contents compared to Earth. The region beyond 2 AU and within the current orbit of Jupiter is suggested to originally consist of isotopically ordinary-chondrite (OC) like material which is moderately volatile rich and oxidized, whereas the terrestrial planets region are likely to be initially consist of volatile poor and reduced enstatite-chondrite (EC) like material (Carlson et al., 2018; Fischer-Gödde and Kleine, 2017; Render et al., 2017). Our simulation indicates embryos in the Mercury and Venus region after the gas disc dissipates could have a higher fraction of OC-like material than Earth. Alternatively, if an isotopic gradient existed within the inner Solar system before the planetesimals formation (Mah and Brasser, 2021; Yamakawa et al., 2010), such gradient would be erased after the dissipation of the gas disc and the formation of embryos due to the strong mixing effect.

Compared to embryos in the Mercury and Venus regions, embryos in Earth's region typically have an accretion zone in between 1 to 2 AU, which is generally less wide than those in the Mercury and Venus region. In our simulations, the sweeping secular resonance does not bring a large amount of material to the Earth region. Table 2 shows the mass percentage of embryos in the Mercury, Venus, Earth and Mars regions that originated from different distances in the disc. In both the classical and depleted disc models, the percentage of material originating from beyond 2 AU that is deposited in the inner solar system is twice as high for the Venus region embryos compared to Earth region embryos. This indicates that the sweeping secular resonance is less effective at transporting material from the asteroid belt to the Earth region in our simulation. On the other hand, in the classical model, Mercury and Venus consist of one-third to nearly half of their mass from asteroid belt material, which is surprisingly high.

Our results may also infer a similar composition between embryos in the Earth and Mars region. The accretion zones of embryos near Mars overlap with Earth's according to figure 7. Data from table 2 also suggests a similar bulk composition between them. The percentage of mass originating from 1 to 2 AU that end in Earth and Mars region's embryos are both over 70%, indicating a likely similar composition between them. Recent isotopic data suggest differences in bulk composition between Earth and Mars (Brasser et al., 2018; Dauphas, 2017; Lodders and Fegley, 1997; Sanloup et al., 1999; Tang and Dauphas, 2014). A higher resolution accretion zone analysis together with full terrestrial planet simulations of over 100 Myr is required to further understand the outcome.

Although results from the classical model and depleted disc model are similar, there are still recognisable differences. In the classical model embryos in the Mercury and Venus regions accrete a higher percentage of material beyond 2 AU than those in the depleted disc model (table 2). This difference is more than twice for Mercury. This is because of the lower initial solid surface density in the outer region of the disc and (in some cases) a higher surface density in the inner region than the MMSN in the depleted disc model (see figure 1). Hence the total mass flux of materials beyond 2 AU mixing into the inner region by the sweeping secular resonance is lower in the depleted disc model. This also explains why in the classical model there are more embryos in Mercury and Venus regions with $a_{\text{weight}} > 2$ AU and nearly all embryos in the Mars region with $a_{\text{weight}} > 1.5$ AU.

We have also performed a low resolution test simulation to examine whether including material beyond 3 AU would affect our final results. We found that > 60% of the material beyond 3 AU are ejected due to the strong gravitational perturbation of Jupiter, and only < 40 % of this material are delivered to the inner Solar System, mostly to the Mercury-Venus region. In contrast only ~3% planetesimals originated from 2 to 3 AU are ejected from the system, and the remaining ~97 % are mostly transported to the inner Solar System. Hence, the material beyond 3 AU only has a relatively minor effect on embryos' composition compared to those within 3 AU. This justify our choice of selecting 3 AU as the outer edge of our solid disc.

It should be mentioned that the results of the embryos accretion could be different if we assume that the giant planets also feel the gas potential of the disc. In that case the eigenfrequency $g_5$ would be different and hence $\nu_5$ sweeps through the asteroid region at a different time. This could potentially affect the embryo final composition by implanting materials > 2 AU to a further region of the disc instead of the current Mercury-Venus region. Further simulation needs to be carried out in order to verify it.

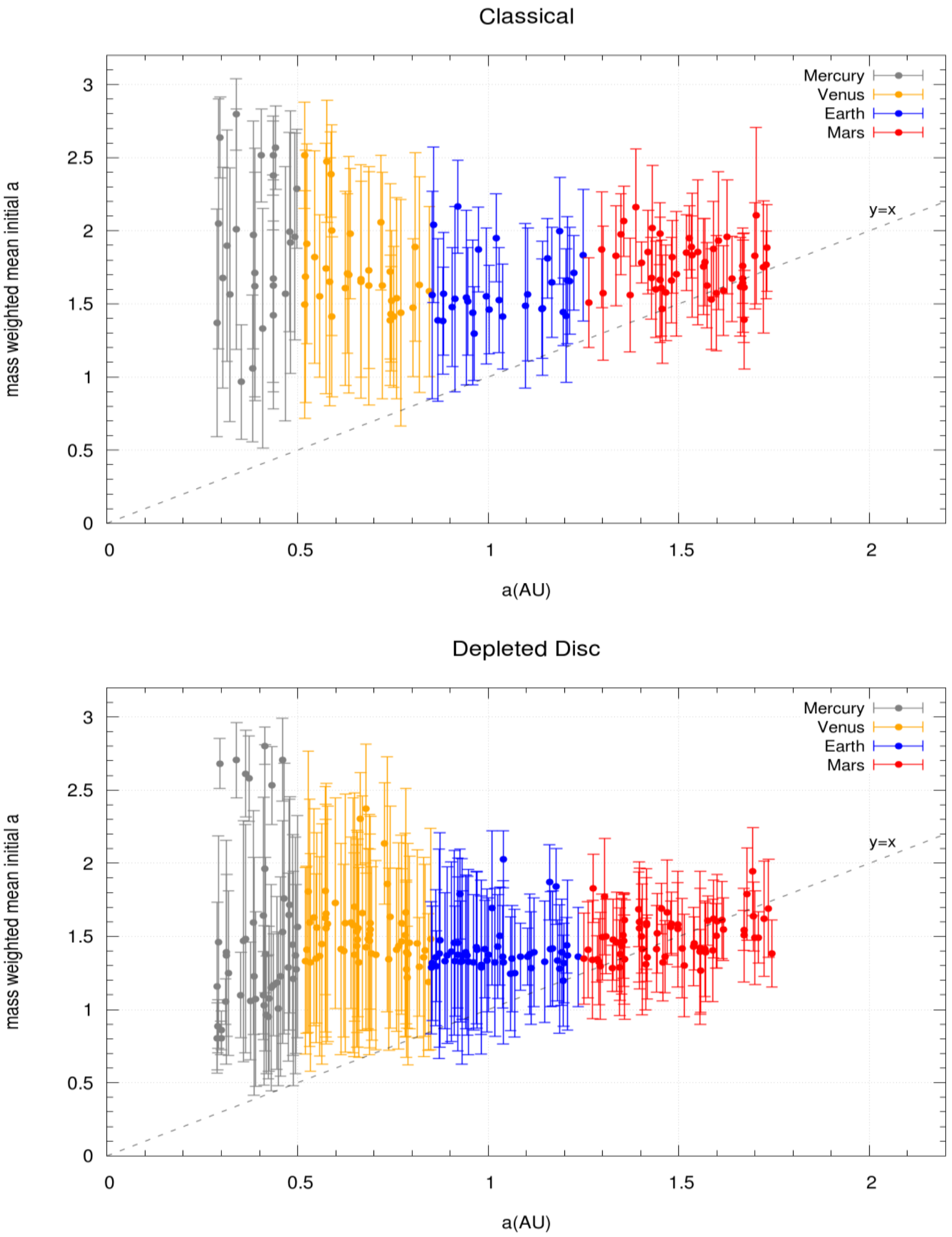

Figure 7 - Mass weighted mean initial semi-major axis ($a_{\mathrm{weight}}$, equation 15) versus the final semi-major axis of Mercury (grey), Venus (orange), Earth (blue) and Mars (red) embryos in all simulations of the classical EJS model (CL1+CL1_LR) (upper panel) and the depleted disc (DP1 to DP4 + DP1_LR) (lower panel) with $r$ = 350 km for initial planetesimals and gas disc $\tau_{decay}$ = 2 Myr. Only the embryos survive at 10 Myr are plotted**.** The error bars represent $\sigma_{\mathrm{weight}}$ (equation 16), which is the 1$\sigma$ standard deviation of $a_{\mathrm{weight}}$. The accretion zone of each embryo can be represented by $a_{\mathrm{weight}} \pm \sigma_{\mathrm{weight}}$.

Table 2 - Percentage of mass originated from a < 1 AU , 1 AU < a < 2 AU and a > 2 AU for Mercury, Venus, Earth and Mars embryos in the classical (CL1+CL1_LR) and depleted disc simulation (DP1 - DP4 + DP1_LR) with EJS and gas disc $\tau_{decay}$ = 2 Myr.

| | Classical (EJS) | | | Depleted Disc | | |
|---|---|---|---|---|---|---|
| | a < 1 AU | 1 AU < a < 2 AU | a > 2 AU | a < 1 AU | 1 AU < a < 2 AU | a > 2 AU |
| Mercury | 24.5 % | 29.9 % | 45.6 % | 38.8 % | 37.1 % | 24.0 % |
| Venus | 14.7 % | 49.6 % | 35.9 % | 23.2 % | 51.0 % | 25.7 % |
| Earth | 6.2 % | 76.4 % | 17.6 % | 13.6 % | 75.6 % | 10.8 % |
| Mars | 1.3 % | 79.0 % | 19.8 % | 4.9 % | 88.9 % | 6.2 % |

## 3.3. Growth timescale of embryos

We observe strong inside-out growth trends for embryos in figure 2 and figure 4. In this section, we present the timescale for the embryos' growth, focusing more on embryos in the region of Mars since it is the only planet other than Earth for which we have isotopic chronological data constraining its formation timescale. The high resolution and the low resolution data are presented separately in this section since their results are slightly different. In this section, we first present the average time for embryos to reach a lunar mass in each region (section 3.3.1). Second, we discuss the current constraints for Venus' and Earth's embryos inferred from atmospheric measurement and samples (section 3.3.2). Third, we focus on comparing the formation timescale of Mars' region embryos with the Hf-W chronology (section 3.3.3). Finally, we present the statistical analysis on the collisional history of Mars' region embryos and discuss the possibility of a younger Mars inferred from Hf-W chronology (section 3.3.4).

### *3.3.1 Time for embryo reaching lunar mass*

Figure 8 depicts the mean time for any embryos that survive to 10 Myr to reach a lunar mass in each region in the low resolution (planetesimals with initial $r$ = 800 km) and high resolution (planetesimals with initial $r$ = 350 km) simulations (filled circle); we present the output for the classical and the depleted disc models. The uncertainties are the 5th and the 95th percentile of the data. The mean time for planetesimals to accrete to lunar-mass embryos does not have a large difference between the classical and the depleted disc models; their differences are typically less than a Myr. Differences between data from the low resolutions and high resolution simulations are more obvious. There is no clear positive trend for data in the low resolution simulations and the data cover a wider range, i.e. having larger error bars, compared to the high resolution simulations. The time differences between the fastest- and slowest-growing embryos in each region are generally more than 5 Myr in the low resolution simulation, especially in the Mercury and Venus region where the differences can be close to or even more than 8 Myr. Besides, the Mercury and Venus region embryos are usually formed at a later time in the low resolution data compared to the high resolution data. This is because in the low resolution simulations, the initial masses of the planetesimals are an order of magnitude higher than in the high resolution cases. Even though the encounter rate between planetesimals is expected to be lower in low resolution simulation due to the lower total number of planetesimals, objects near the inner edge of the disc can grow to over a lunar mass in less than 0.1 Myr. These extremely fast growing first generation embryos, however, do not survive to the end of the simulation since the density of the gas disc is still high at the early stage. This leads to fast Type-I migration and the embryos are lost due to inward migration to the Sun. In contrast, the first embryos to form at the inner edge take 0.2 to 0.3 Myr to form in the high resolution simulation, which is slightly lower than in the high resolution cases. More fast-forming embryos are retained in the Mercury and Venus region until the end of simulation in the high resolution cases. Another explanation of the slower formation rate of embryos in the Mercury and Venus region of the low resolution simulation is the sweeping secular resonance bringing embryos or proto-embryos formed in the asteroid belt region to the inner Solar system. Slow-forming embryos form in the asteroid belt that are brought to the Mercury and Venus region can only be found in the low resolution simulations.

In the Earth and Mars region, the situation is different from the Mercury and Venus region since here the growth rate is much slower and hence embryos form in the outer region do not reach the Sun due to Type-I migration. The mean time of reaching a lunar mass in Earth's region is ~3 to 4 Myr. Embryos in the Mars region, on average, take at least a Myr longer to grow than those in the Earth region. In the classical model of the high resolution case, embryos in the Mars region generally take 3 Myr longer to grow than near the Earth. Comparing the classical model's data between high and low resolution indicates that Mars forms quicker in the low resolution cases, with the mean, 5th and 95th percentile are more than 1 Myr shorter than the corresponding high resolution data.

We further present the mean time for the largest embryos in each region of each run to reach a lunar mass in figure 8 (triangular points). These largest embryos could potentially be the seeds of the final planets. In general, both models result in similar timing. The largest embryos form quicker than most of the survived embryos in the same region, as shown by the lower mean for the largest embryos compared to that of all embryos (triangle vs circle for the same colour). The largest embryos in the Mercury-Venus region reach a lunar mass at ~1 to 2 Myr. Those in the Earth region take ~1 to 4 Myr. The Mars region shows significant differences between the high resolution and low resolution simulations, which is not observed for the other regions. The mean time for the largest Mars region embryos to reach a lunar mass is ~2 Myr in the low resolution simulations, whereas the mean time increases to ~4 Myr in the high resolution simulations. This may suggest a speeding up for Mars' formation if we assume that Mars formed mainly from larger planetesimals. We will study Mars' growth timescale in more detail in Section 3.3.3.

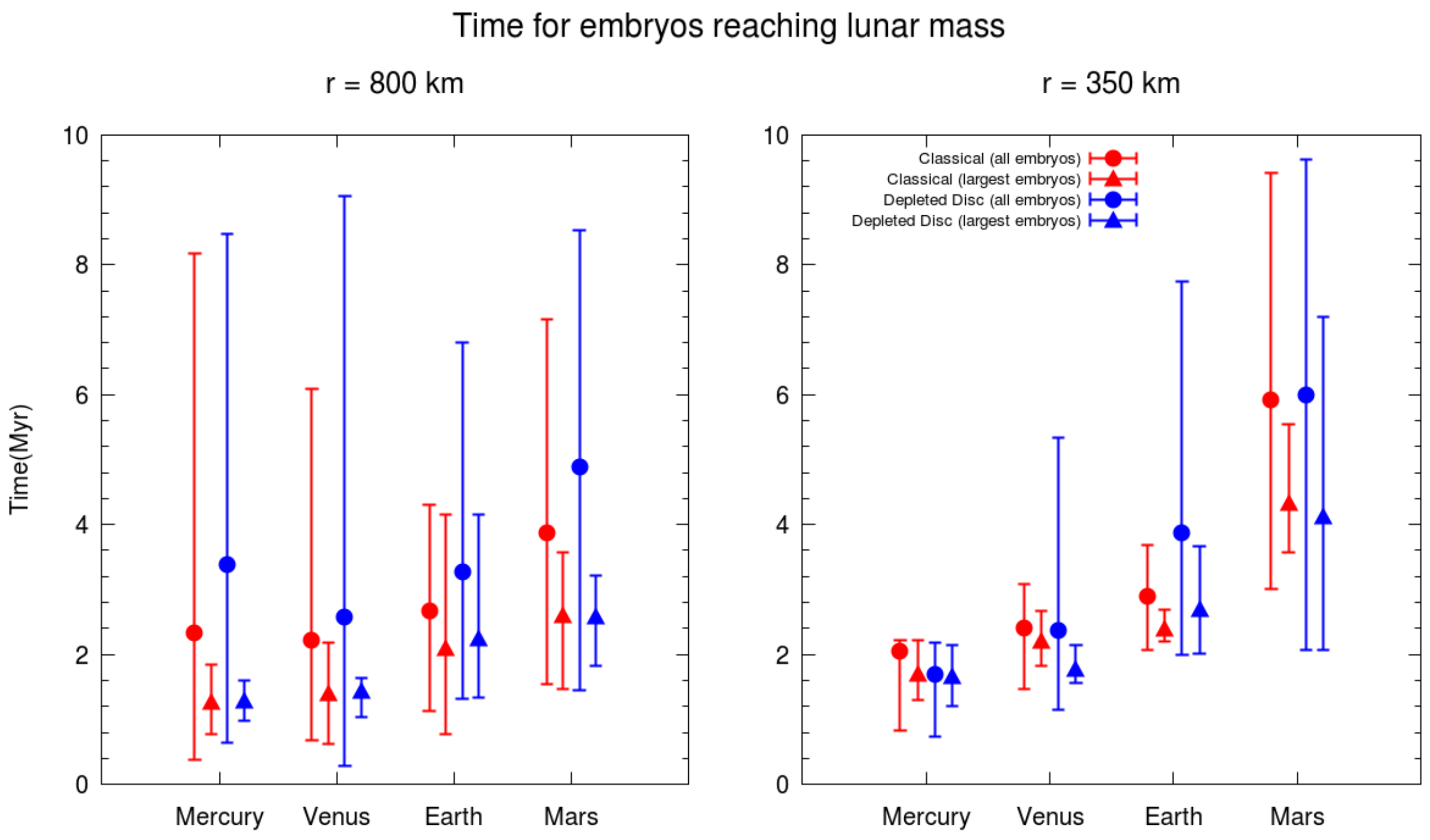

Figure 8 - Mean time for planetesimals to form lunar mass embryos in the Mercury, Venus, Earth and Mars regions. Data is shown for the classical model and the depleted disc models low resolution (*r* = 800 km; left panel) and high resolution (*r* = 350 km; right panel) in a gas disc with $\tau_{decay}$ = 2 Myr. Circular points are the average time of all embryos in that region, whereas triangular points are for only the largest embryos in each region. The error bars represent the 5th and 95th percentile of the data.

### *3.3.2 Possible constraints for Venus' and Earth's growth*

Currently we do not have samples from Mercury and thus no constraint can be made on its formation timescale. Although known samples from Venus also do not exist currently, a recent study applies the atmospheric noble gases isotopic ratio (e.g. $^{22}Ne/^{20}Ne$, $^{36}Ar/^{38}Ar$) measured by the Venera missions (Istomin et al., 1980, 1983) and the Pioneer Venus orbiter (Donahue, 1986; Hoffman et al., 1980), and K/U ratio by the Vega and Venera missions (Davis, 2005) to constrain the upper limit of Venus's growth timescale (Lammer et al., 2020). A fast-growing protoplanet which grew to > 0.5 $M_{\mathrm{Earth}}$ within the lifetime of the Solar nebula could possibly accrete a primordial hydrogen atmosphere with unfractionated Solar isotopic ratios and volatile abundances (e.g. Ikoma and Genda, 2006; Stökl et al., 2016, 2015). Such an early hydrogen atmosphere was then lost through hydrodynamic escape due to the strong Solar extreme-ultraviolet radiation (Johnstone et al., 2015; Lammer et al., 2014; Tu et al., 2015). This process could fractionationate atmospheric noble gases isotopes and alter volatile abundances (Hunten et al., 1987; Odert et al., 2018; Zahnle and Kasting, 1986), leading to the current measured ratio of the planet. Based on this hydrodynamic escape theory, it is suggested that Venus could have grown to its full mass within the solar nebular lifetime by matching its noble gas isotopic ratios and K/U ratio with a hydrodynamic escape atmospheric model (Lammer et al., 2020). Such extremely rapid formation of Venus is not reproduced in our simulations. The largest object from our 10 Myr simulations is only about 2 times more massive than Mars. However, it should be mentioned that the constraints made by noble gas isotopes and volatile abundances from Lammer et al. (2020) only provide a lower limit on the growth timescale and thus do not rule out a slower forming Venus similar to the results from our simulation.

Atmospheric noble gases and volatile constraints can also be applied to Earth. The proto-Earth is proposed to reach 0.53-0.58 $M_{\text{Earth}}$ within the lifetime of the Solar nebula (Lammer et al., 2020). Similar to Venus, this is an upper limit estimation and our study does not reproduce such a rapid formation rate for embryos in the Earth region. Nevertheless, a rapid formation of Earth has been suggested from the $^{48}$Ca (Schiller et al., 2018) and $^{54}$Fe composition of the bulk silicate Earth (Schiller et al., 2020). In their scenario, proto-Earth was originally formed from inner disc material having a Ureilite-like (a type of achondrite with the lowest-known $^{48}$Ca composition) and then grew by continued accretion of inward drifting mm-sized CI-like (a type of carbonaceous chondrite enriched in $^{48}$Ca compared to inner Solar system materials) chondrules due to gas drag (Johansen et al., 2015), instead of the accretion of km-sized planetesimals and embryos. The formation of proto-Earth has to be very rapid (within the lifetime of the Solar gas disc) in order for Earth to accrete a sufficient amount of inward drifting CI-like chondrules within the lifetime. The time suggested by Schiller et al. (2018) for the in-situ formation of the proto-Earth is 0.9 $M_{\text{Earth}}$ within the gas-disc lifetime (Johansen et al., 2015) based on the linear correlation between the logarithmic masses and ages of the inner solar system bodies with their measured $^{48}$Ca composition (See Figure 2 of Schiller et al. (2018)). Although this scenario is also able to explain the rapid formation of Mars, from the dynamical perspective it has difficulties in explaining why the collisions between planetesimals and embryos were not involved in the formation of Earth (Morbidelli, 2018) and the sudden termination of Mars' and achondrites parent bodies' accretion of inward drifting CI-like chondrules. The proposed rapid formation of Jupiter (Kruijer et al., 2017a) or the formation of a pressure maximum in the disc (Brasser and Mojzsis, 2020) have been suggested to shut off the inward drifting chondules flux into the inner Solar system, providing another obstacle for this scenario (Morbidelli, 2018).

A more widely adopted approach of determining the core-formation and thus accretion timescale of planetary objects is the Hf-W chronology. By measuring the Hf/W ratio and the excess of $^{182}$W from the decay of $^{182}$Hf (half life = 8.9 Myr) in the samples from Earth's mantle, Yin et al. (2002) and Kleine et al. (2009) deduces that Earth's core formation and accretion take ~30 Myr to more than 100 Myr to completed, much longer than the aboved chondrule accretion scenario of Schiller et al. (2018). This timescale is more consistent with our simulation and other conventional N-body simulation results (e.g. Brasser et al., 2016; O'Brien et al., 2006; Raymond et al., 2009). The suggested rapid formation of Earth merits further investigation.

### *3.3.3 Growth of Mars: N-body results vs Hf-W chronology*

While the current early evolution of Venus and Earth remains elusive, Mars, on the other hand, has less debate on its formation timescale. We thus make a complete comparison between our N-body results with the meteoritic-inferred formation timescale of Mars. It is suggested that Mars was formed much before Earth. Based on the Hf-W ratio and $^{182}$W measurement on the shergottites, a group of martian meteorites that likely represent the bulk composition of the martian mantle (Kruijer et al., 2017b), Dauphas and Pourmand (2011) proposed that Mars completed its formation within 10 Myr after formation of CAIs. The black solid line in figure 9 is their proposed analytic formation path for Mars by assuming the semi-oligarchic growth of Chambers (2006):

$$M_{\mathrm{Mars}}(t)/M_{\mathrm{Mars}} = \tanh(\mathrm{t}/\tau_{\mathrm{grow}}) \tag{17}$$

According to Dauphas and Pourmand (2011) $\tau_{grow} = 1.8^{+0.9}_{-1}$ Myr from Hf-W chronology; the uncertainties represent the 95% confidence interval. Another study based on a similar approach but with the $^{60}Fe$-$^{60}Ni$ chronological system (half life = 2.62 Myr) obtained similar results with $\tau_{\mathrm{grow}} = 1.9^{+1.7}_{-0.8}$ Myr (Tang and Dauphas, 2014). These rapid formation time scales suggest that Mars had finished accreting nearly half of its mass within 2 Myr after the formation of the Solar system.

The red lines in figure 9 are the growth curves of embryos of Mars within the first 10 Myr in our low and high resolution of classical and depleted disc simulation, respectively. All growth curves of the Mars region embryos from our N-body simulations are below the black line and mostly do not overlap with the yellow-green regions, indicating that our simulations have difficulty in reproducing the fast growth of Mars. None of the Mars region embryos reach close to 0.5 $M_{\mathrm{Mars}}$ within 2 Myr, as suggested by Dauphas and Pourmand (2011) and Tang & Dauphas (2014). Two of the fastest growth of Mars analogues are in the low resolution ($r$ = 800 km) simulations of the classical case, both reaching ~0.6 $M_{\mathrm{Mars}}$ by 5 Myr and 1 $M_{\mathrm{Mars}}$ within 10 Myr, indicating a possibility of forming Mars quick enough if Mars mainly form from larger planetesimals in a solid disc without depletion in the initial solid surface density with respect to the MMSN (i.e. classical model instead of depleted disc model). Only one Mars region embryo has reached 1 $M_{\mathrm{Mars}}$ within 10 Myr in the high resolution simulations. This embryo grows over 1 $M_{\mathrm{Mars}}$ because of a giant impact at 9 Myr which nearly doubles its mass. However, it still has an initially slow growth and does not reach 0.5 $M_{\mathrm{Mars}}$ within 5 Myr. This could suggest that for Mars to finish growing within 10 Myr, a giant impact near the end of its accretion, or several such events during its early accretion, may be required.

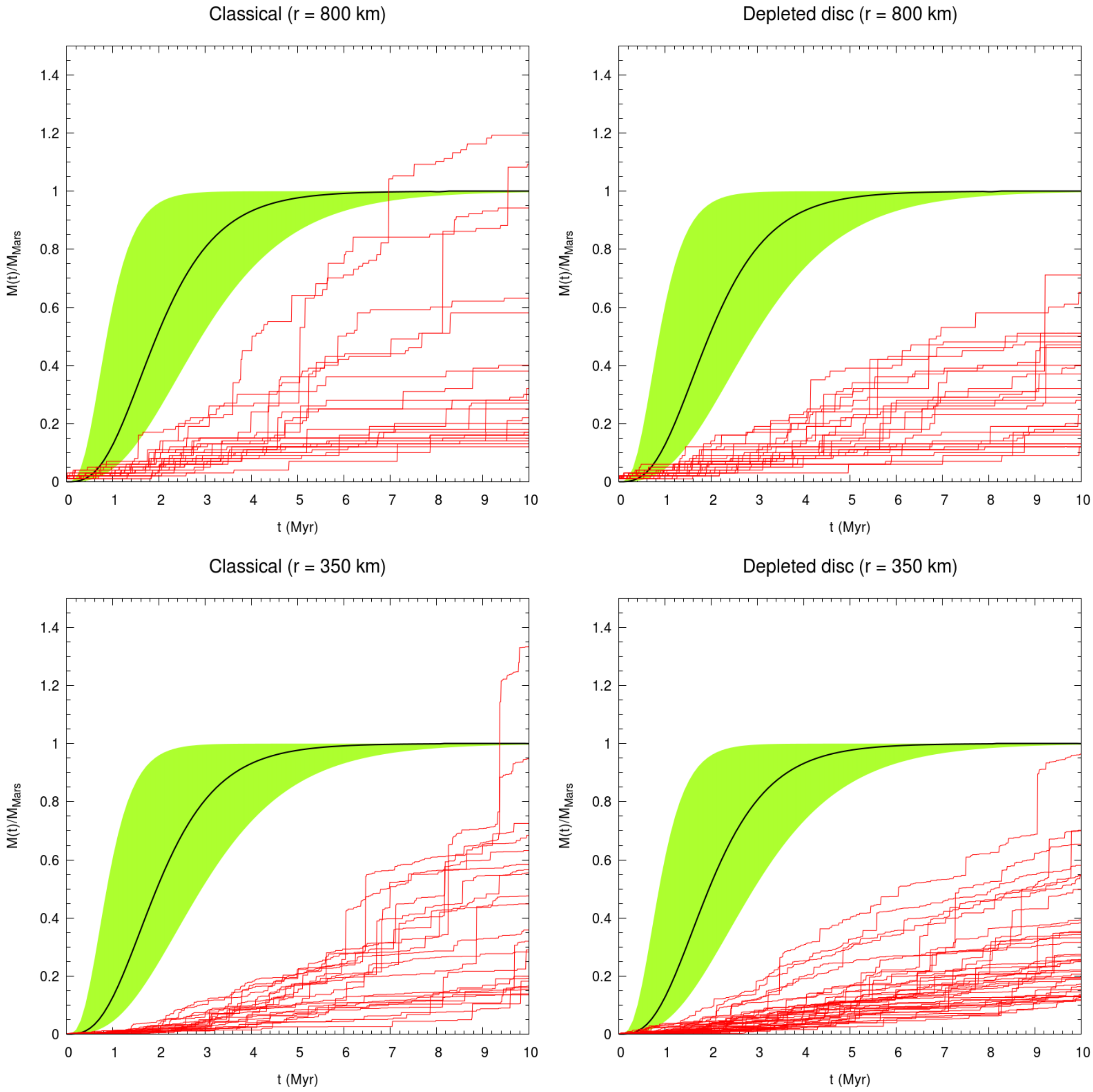

Figure 9 - Comparison between the growth curves of Mars from N-body simulation and the suggested theoretical growth model based on meteoritic data. The black line is the growth curve suggested by Dauphas and Pourmand (2011) based on the Hf-W chronology on Shergottites. The yellow-green band represents the 95% confidence interval of their model. The red lines are the growth of Mars' embryos from our N-body simulation in the classical model (left panels) and the depleted disc model (right panels) with $\tau_{decay}$ = 2 Myr for the gas disc. The initial radii of planetesimals are 800 km (upper panels) and 350 km (lower panels).

In order to further study how changing the resolution (i.e. initial planetesimals' radii and number) affects the growth of planetesimals near the Mars region, we simulate the growth of embryos within an annulus from 1.45 to 1.55 AU with Jupiter and Saturn on their current orbits. The planetesimals have initial radii of $r$ = 125 or 500 km. Table 1 shows the initial condition of the annulus simulations (Annu_125km and Annu_500km). The solid surface density is based on the MMSN. The total mass in this annulus is ~0.12 $M_{\text{Earth}}$, which is about 1 $M_{\text{Mars}}$. Given the low mass within the annulus, we do not expect these annulus simulations to form objects close to 1 $M_{\text{Mars}}$ within 10 Myr. Hence we are not comparing the annulus simulation with other full disc simulations, but only amongst annulus simulations. Figure 10 shows the growth evolution of the largest objects within each of the annuli for 10 Myr. All the objects have barely reached a lunar mass and are still more than 10 times less massive than the current Mars. Even though collisions are more frequent in the simulation with smaller planetesimals case and the two objects converge to a similar mass during the first 10 Myr, the largest object in the simulation with $r$ = 125 km remains less massive than the one begin as a larger planetesimals during most of the time in the first 10 Myr. Our results suggest that it could be more likely to match the fast growth of Mars within the first 5 Myr when the initial diameter of the planetesimals is larger.

The Mars embryo growth timescales in our simulations are roughly consistent with those in Walsh & Levison (2019) that used a different methodology. Morishima et al. (2013) also applied high resolution N-body simulations to study Mars' growth within an annulus at ~1.4 to 1.6 AU and compared their results to the Hf-W chronology. Similar to most of our results, their simulations fail to form an embryo reaching 1 $M_{\text{Mars}}$ at 14 Myr, even though they begin with a solid surface density higher than the MMSN. The largest object in their simulations reaches only about 80 to 85% of $M_{\text{Mars}}$. Their slow growth of Mars can be attributed to the lack of material scattered in from outside the Mars region. Different from Morishima et al. (2013), some of our simulations form embryos with mass > 1 $M_{\text{Mars}}$ because our simulations are performed for a full disc. Here Mars can accrete material from everywhere in the disc, instead of from only a small annulus. Materials from other regions of the disc are likely required to form Mars more quickly.

However, our results contradict the proposed theoretical idea of forming Mars fast within 10 Myr from small planetesimals ($r$ < 10 km) (Kobayashi et al., 2010; Kobayashi and Dauphas, 2013). There are several reasons to explain the discrepancy between our results and previous analytic theory. First of all, our N-body simulations do not include collisional fragmentation. Smaller planetesimals are more susceptible to collisional fragmentation than larger ones. A fraction of fragments' masses could be damped down to cold orbits by the gas drag and this increases the collisional cross section of the growing embryos and could potentially speed up the growth of embryos if the solid disc was initially composed of planetesimals with $r$ < 10 km (Kobayashi and Dauphas, 2013). Another reason for the discrepancy is the assumed solid surface density. We assumed the MMSN in our classical model simulations, whereas Kobayashi and Dauphas (2013) proposed a solid surface density more than 2 times higher than the MMSN is able to form Mars within 10 Myr from planetesimals with $r$ < 10 km. The higher solid surface density would not lead to overshooting Mars (e.g. Chambers, 2001) because some fragments generated by collisions between small planetesimals could be lost to the Sun due to gas drag. However, the current N-body method is still unable to simulate such a high resolution system with millions of particles because of the limit of computational power. With their $N$-body code, Walsh & Levison (2019) have shown that initial planetesimals with $r$ = 30 km together with fragmentation still have difficulty reaching 50% $M_{\text{Mars}}$ within 5 Myr for their Mars region embryos. More statistics from high resolution $N$-body simulations including fragmentation are needed to further investigate this problem and how fragmentation affects the growth rate (Quintana et al., 2016).

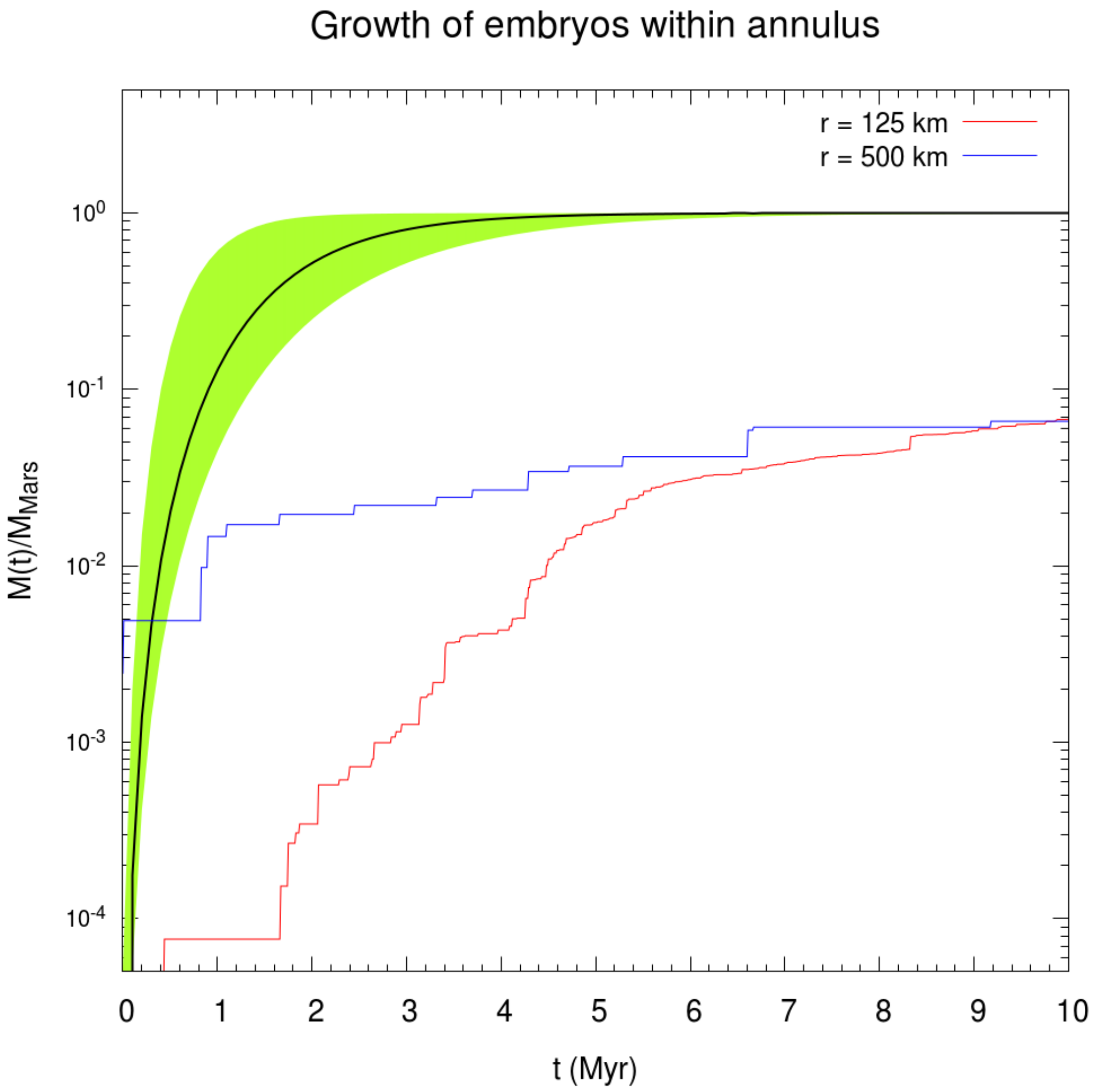

Figure 10 - Same as figure 9, but for the largest object within the annulus at 10 Myr. The initial radii of the planetesimals are either 125 or 500 km (Annu_125km and Annu_500km, Table 1).

### *3.3.4 Mars' collisional growth: possible uncertainties of Hf-W chronology?*

The result of forming Mars within 10 Myr from Hf-W chronology has been challenged by a recent study combining SPH simulations and Martian W isotopic data (Marchi et al., 2020). It has been shown through SPH simulations that the incomplete mixing of the impactor's material in proto-Mars' mantle could lead to heterogeneity of $^{182}$W in the martian mantle. Indeed, different types of martian SNC meteorites, such as the nakhlites and ALH84001 (Foley et al., 2005; Kruijer et al., 2017b; Lee and Halliday, 1997) have different measured $^{182}$W isotope data. Hence, Marchi et al. (2020) suggested that if we assume a lower $\varepsilon^{182}$W (the $\varepsilon$ -notation denotes deviations in parts-per-ten thousand normalised to another standardised isotopic ratio of the same element), such as the one measured from ALH84001, as the representative of bulk martian mantle composition, the inferred formation time scale could be up to 15 Myr for Mars. However, this longer time scale is unlikely to be correct because it has been shown that the enriched shergottites instead of the naktitles or ALH84001 are more likely to be representative of the bulk martian mantle based on their correlated $\varepsilon^{182}$W–$\varepsilon^{142}$Nd variations (Kruijer et al., 2017b) combined with an estimate of the bulk martian mantle $\varepsilon^{142}$Nd (Mezger et al., 2013).

The uncertainty of Mars' formation timescale inferred from Hf-W chronology can also be caused by incomplete equilibration between the impactor's core and proto-Mars' mantle. The equilibration factor $k$ is defined as the fraction of the impactor's core not merged directly with the target's core (Nimmo et al., 2010). Therefore, $k = 1$ means complete chemical mixing of the impactor's core into the target's mantle, whereas $k = 0$ represents core-merging between impactor and target. It has been shown that having lower $k$ throughout the whole accretion of Earth would lead to a longer inferred growth timescale $\tau_{grow}$ (> 100 Myr) than higher $k$ (< 20 Myr) (Rudge et al., 2010). The estimation of $\tau_{grow}$ ~ 2 Myr by Dauphas and Pourmand (2011)and Tang and Dauphas(2014) for Mars assumes $k$ ~ 1 (complete equilibration) for the whole duration of Mars' accretion. They justified this based on two reasons. First, Mars mainly accreted small planetesimals (10 to 100 km across) throughout its formation. Second, Mars formed early when short lived radioactive isotopes such as $^{26}$Al are still active, and thus Mars likely developed a global magma ocean on its surface (Bouvier et al., 2018; Debaille et al., 2007; Kruijer et al., 2020; Morishima et al. 2013). Together with the possibility of the planetesimals also being molten and striking proto-Mars' magma ocean, complete equilibration between planetesimals and the proto-Mars could easily be achieved (Morishima et al. 2013).

Contrary to the former assumption, our results show that Mars' growth also involves large impactors (dubbed "giant impact type accretion", i.e. $M_{\text{impactor}}/M_{\text{target}} > 0.1$, where $M_{\text{impactor}}$ and $M_{target}$ are impactor's and target's mass, respectively) instead of only accreting planetesimals. This impactor to target ratio for giant impacts is arbitrary and is simply based on the expected ratio for the canonical Moon-forming collision (Canup & Asphaug 2001)**.** Table 3 shows the percentage of the total mass delivered by impactors > 10% of the mass of the growing Mars at the time of collision. In the classical model, nearly half of the mass of the embryos in Mars region are delivered by giant impact type accretion, whereas this percentage is slightly lower to ~30 to 40 % in the depleted disc model. Besides, impactors with mass greater than lunar mass delivered > 30% of mass to the Mars region embryos (i.e. an embryo-embryo collision) in the classical model, whereas only < 20 % of mass are delivered by such massive impactors in the depleted disc model.

The collision frequency between the forming Mars region embryos and large impactor relative to their size at the time of collision depends on the initial assumed size of planetesimals. Figure 11 shows the cumulative distribution of impactor's to target's ratio of collisions experienced by embryos in Mars' region. The results of the classical and the depleted disc models are similar. About 30% of collisions are in the giant impact regime in the low resolution simulation, whereas this percentage decreases to ~ 5 % when the initial planetesimals' size decreases to $r$ = 350 km. This suggests that if Mars mainly formed from more massive planetesimals', giant impact-type collisions would play a more important role in Mars' formation history hence leading to a higher probability for low-$k$ collisions to occur because it is more difficult to fully mix a large impactor's core completely into proto-Mars' mantle than a small impactor's core (Deguen et al., 2011; Samuel, 2012). Therefore, if Mars mainly formed from accretion of larger planetesimals (e.g. $r$ > 350 km), the Hf-W ages for Mars could possibly be younger than the age inferred from Dauphas and Pourmand (2011) and Tang and Dauphas (2014) (i.e. Mars could be younger than previously thought and thus formed slower; Marchi et al. 2020).

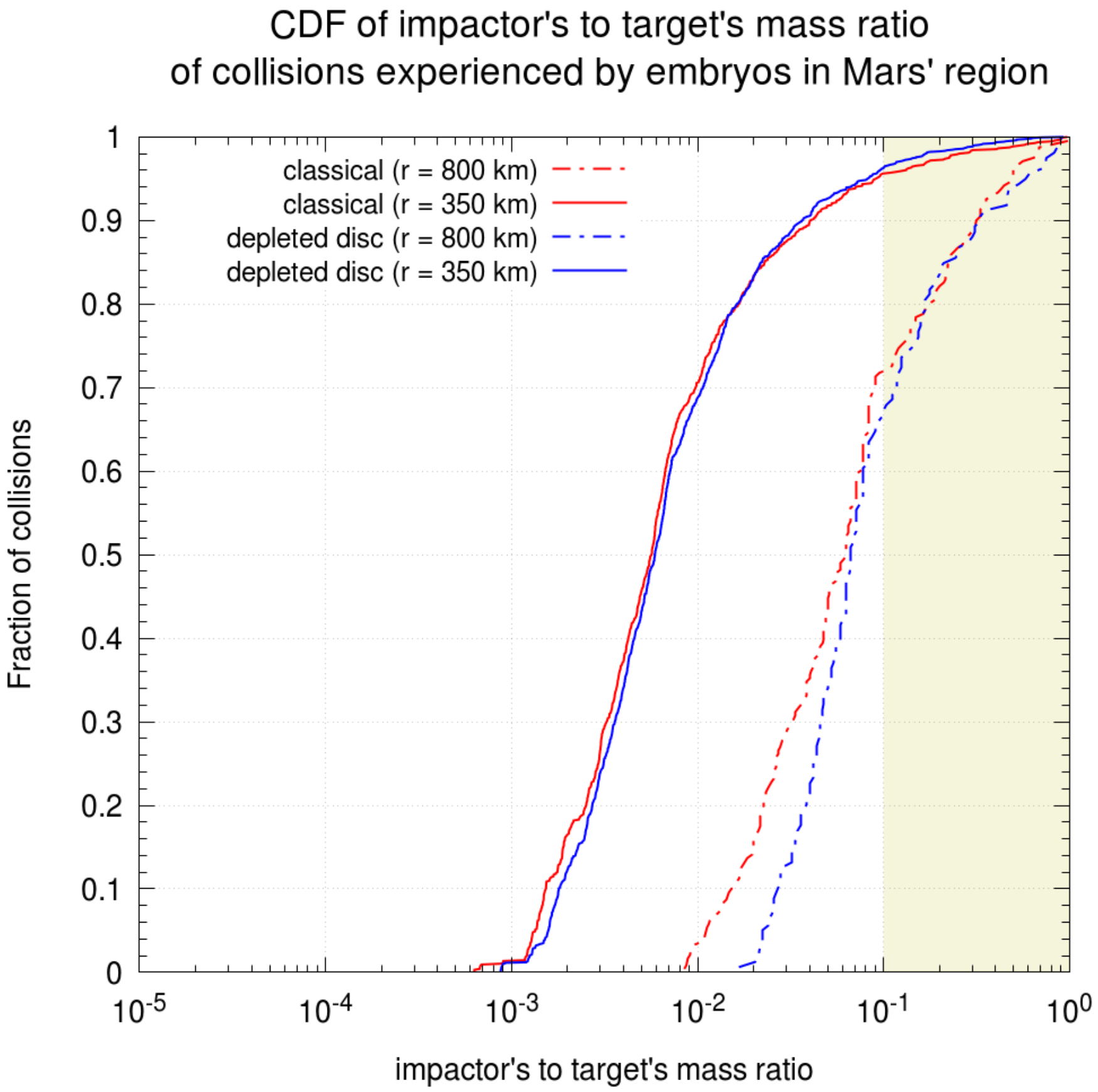

Figure 11 - Cumulative distribution function of the impactor to target ratio of the collisions experienced by embryos in the Mars region at 10 Myr in the classical model including EJS and CJS simulations (red lines) and the depleted disc model (blue lines) with gas disc $\tau_{decay}$ = 2 Myr, with initial radii of planetesimals as 800 km (dashed-dotted line) or 350 km (solid line). The shaded region corresponds to the impactor to target ratio > 0.1, which is regarded as the giant impact regime in our study. Collisions are counted only after the Mars region embryos have reached 1 $M_{\text{Moon}}$.

Table 3 - Total percentage of mass delivered by giant impact type collision ($M_{\text{impactor}}/M_{\text{target}} > 0.1$) and embryo-embryo collision ($M_{\text{impactor}} > 1\ M_{\text{Moon}}$) for embryos in Mars region. The simulations are the classical model including EJS and CJS simulations and the depleted disc model with gas disc $\tau_{decay}$ = 2 Myr, with initial radii of planetesimals as 800 km or 350 km. Collisions are counted only after the Mars region embryos have reached 1 $M_{\text{Moon}}$

| Impactor | Classical | Depleted Disc |
|---|---|---|

| mass ($M_{impactor}$) | $r$ = 800 km | $r$ = 350 km | $r$ = 800 km | $r$ = 350 km |
|---|---|---|---|---|
| > 0.1 $M_{target}$ | 50.4 % | 41.8 % | 40.8 % | 27.4 % |
| > 1 $M_{moon}$ | 31.3 % | 34.4 % | 16.3 % | 10.3 % |

## 4. Changing the gas disc and the initial orbits of the giant planets

In the previous section we presented results of simulations in a gas disc with $\tau_{decay}$ = 2 Myr, a value that was adopted by Walsh & Levison (2019). This long decay time scale means that the gas disc exists for about 10 Myr after the birth of the Sun. Recent meteoritic magnetism data inferred a shorter lifetime, ~3.7 to 3.9 Myr for the protosolar gas disc (Wang et al., 2017). Pb-Pb dating of chondrules also supports a short lifetime of the gas disc, between ~3.3 to 4.5 Myr (Bollard et al., 2017). Therefore, we also perform simulations with a shorter life time of the gas disc ($\tau_{decay}$ = 1 Myr). We also run some simulations without the gas disc. Both sets are intended to study how the gas decay timescale changes the outcome of the simulations. In this section, we show that altering the gas disc lifetime affects embryos' growth (section 4.1) and embryos' accretion zones (section 4.2). In the accretion zone analysis we further demonstrate that changing the initial orbits of the giant planets to being more circular has an even greater impact in embryos' accretion zones.

### 4.1. Effect on embryo growth

#### *4.1.1. Mass evolution*

With a shorter decay timescale we expect Type-I migration to last much shorter and thus fewer embryos should be lost to the Sun. In addition the effect of the secular resonance sweeping should occur earlier, and the whole disc should become more excited at an earlier stage as well. Figure 12 shows the mass versus semi-major axis (left panels) and eccentricity versus semi-major axis of the system at different times for the high resolution EJS classical simulations with $\tau_{decay}$ = 1 Myr (CL1_T1). Like the simulations with $\tau_{decay}$ = 2 Myr lunar-sized embryos emerge after 1 Myr within 1 AU, but do not emerge in the Mars region until ~3 Myr. At this time the asteroid belt region begins to be affected by the sweeping $\nu_5$ secular resonance and thus material in the outer region is transferred to the inner terrestrial planet forming region. Compared to the case with $\tau_{decay}$ = 2 Myr (figure 2), this process begins ~3 Myr earlier when $\tau_{decay}$ = 1 Myr. However, due to the shorter lifetime of the gas disc, the sweeping secular resonance effect stops after ~5 Myr of evolution, regardless of the continued inwarding moving $\nu_5$ location, and material beyond Mars has been mostly depleted. Due to the much earlier transport of material from the outer to the inner region by the sweeping secular resonance the solid surface density of the terrestrial planet region is enhanced earlier and thus embryos in the terrestrial planets region grow faster than in the case with the long-lived gas disc. As a result embryos can grow over 3 $M_{Mars}$ within 10 Myr which did not happen in the previous set of simulations. The growth of some embryos beyond a few times the isolation mass in the terrestrial planet region is due to merging between embryos. Table 4 shows the average mass of embryos in all EJS classical simulations for both gas disc decay times. The average mass of embryos is higher for the short-lived gas disc. The difference between the average mass of embryos can be nearly two times within the high resolution simulations. Nevertheless, the growth of embryos in Venus and Earth region is still not fast enough to match the rapid timescale suggested by the noble gas isotopic composition (Lammer et al., 2020) and $^{48}Ca$ and $^{54}Fe$ isotopic composition (Schiller et al., 2018, 2020) (see Section 3.3 for discussion).

As expected, the mass loss to the Sun due to Type I migration is the lowest in high the resolution simulations with $\tau_{decay}$ = 1 Myr, losing only ~0.03 $M_{\mathrm{Earth}}$ compared to more than 25 times more mass loss in the low resolution simulations with $\tau_{decay}$ = 2 Myr. As explained in the previous section, embryos reach the mass at which Type-I migration sets in much earlier when beginning the simulations with larger planetesimals (i.e. lower resolution). In a long lived gas disc with $\tau_{decay}$ = 2 Myr, about 0.5 to 1 $M_{\mathrm{Earth}}$ of material is lost to the Sun when assuming an initial MMSN solid surface density. In either case the remaining disc mass is about 2.5 to 3 $M_{\mathrm{Earth}}$, which is sufficient for the formation of the current terrestrial planetary system. Our results indicate that the initial mass (or size) of the planetesimals and the lifetime of the gas disc is crucial in determining the mass-distance distribution of the formed embryos.

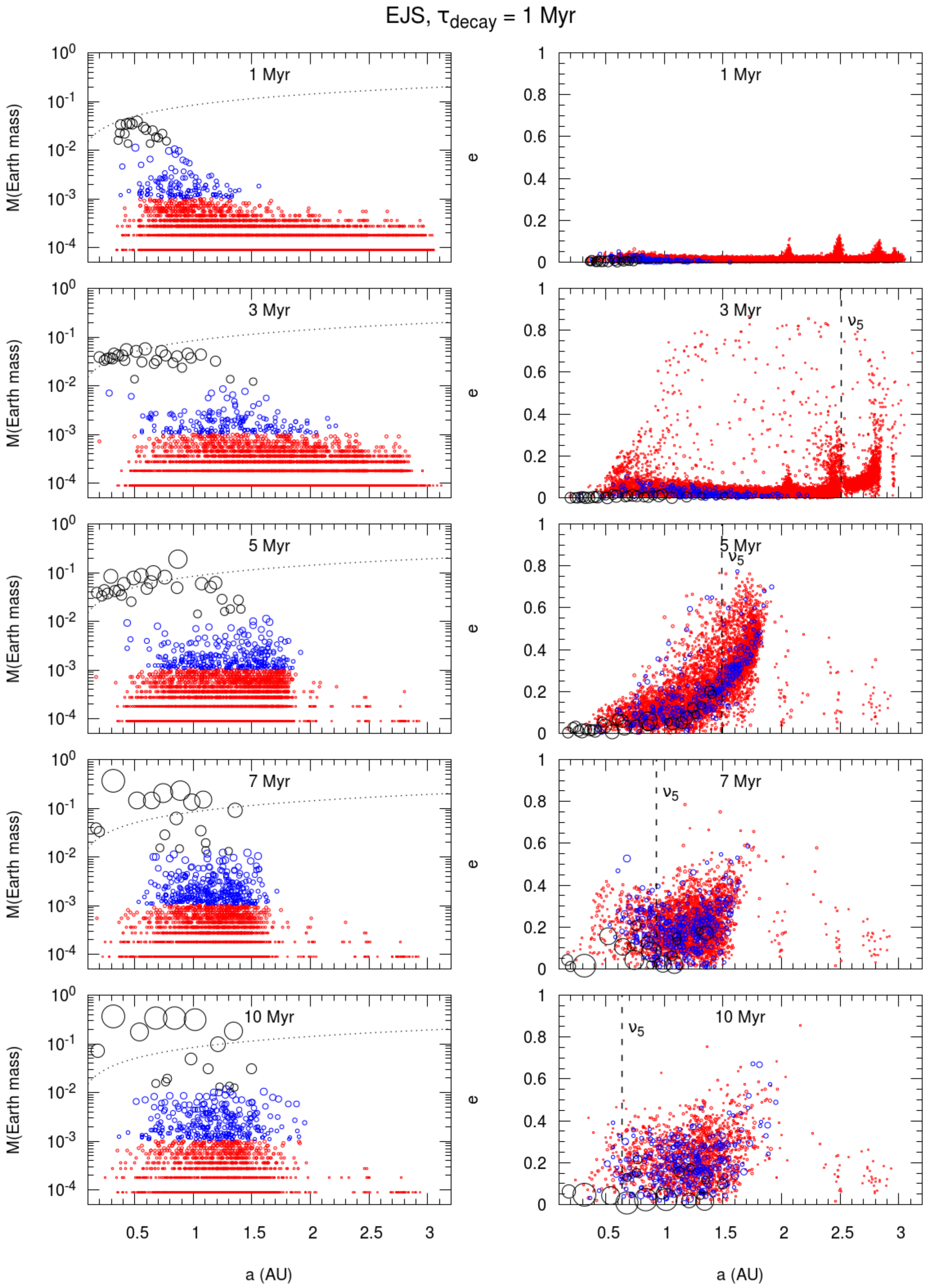

Figure 12 - Mass versus semi-major axis (left panels) and eccentricity versus semi-major axis of planetesimals (red), proto-embryos (blue) and embryos (black) at 1, 3, 5, 7 and 10 Myr in the simulation of the classical EJS simulation with gas disc $\tau_{decay}$ = 1 Myr (CL1_T1). The initial radii of planetesimals are 350 km. The dotted line of the left panels represents the isolation mass of the embryo and dashed line of the right panels represents the location of the $\nu_5$ secular resonance computed from the appendix of Nagasawa et al. (2000).

Table 4 - Average mass of embryos, mass loss to the Sun per simulation and leftover mass of the disc per simulation. We list data for the EJS classical model low resolution ($r$ = 800 km) and high resolution ($r$ = 350 km) simulation. The dissipation time scale of the gas disc $\tau_{decay}$ to 1 Myr or 2 Myr. The total initial mass of planetesimals is ~3.38 $M_{\text{Earth}}$. The average embryo mass and mass left in the disc is listed at 10 Myr.

| | EJS ($r$ = 800 km) | | EJS ($r$ = 350 km) | |
|---|---|---|---|---|
| | $\tau_{decay}$ = 1 Myr | $\tau_{decay}$ = 2 Myr | $\tau_{decay}$ = 1 Myr | $\tau_{decay}$ = 2 Myr |
| Average embryo mass ($M_{\text{Earth}}$) | 0.09 ± 0.09 | 0.07 ± 0.05 | 0.14 ± 0.16 | 0.07 ± 0.04 |
| mass loss to Sun per simulation ($M_{\text{Earth}}$) | 0.12 ± 0.04 | 0.81 ± 0.13 | 0.05 ± 0.02 | 0.54 ± 0.03 |
| Total leftover mass in the disc per simulation ($M_{\text{Earth}}$) | 3.19 ± 0.07 | 2.46 ± 0.13 | 3.31 ± 0.03 | 2.80 ± 0.03 |

#### *4.1.2. Embryos' growth timescale*

In figure 12 and table 4, we show that it is possible to speed up the formation of the embryos and to form larger embryos within 10 Myr when assuming a shorter dissipation time scale for the gas disc. In the previous section, we also showed that it is difficult to explain the rapid formation of Mars when assuming $\tau_{decay}$ = 2 Myr. Here, we show that assuming a short disc lifetime with $\tau_{decay}$ = 1 Myr could possibly speed up the growth rate of embryos at Mars region and explain Mars' fast growth inferred from Hf-W chronology. Figure 13 shows the growth tracks of all embryos in the Mars region for 10 Myr assuming $\tau_{decay}$ = 1 Myr for the classical EJS simulation (left panels). Even though we still found most embryos in the Mars region fail to reach 1 $M_{\text{Mars}}$ within 10 Myr, the low resolution cases contain at five embryos that reach 1 $M_{\text{mars}}$ and two of them have growth rates matching the growth of Mars as suggested by Dauphas and Pourmand (2011). The growth curves of these embryos overlap with the yellow-green region in figure 12. Compared to the low resolution simulations with $\tau_{decay}$ = 2 Myr (upper left of figure 9), in which the growth of Mars region embryos are all below the black line and the yellow-green region, it clearly indicates that the fast decay of the gas disc has a higher possibility of yielding faster growth of Mars region embryos. This is explained by the earlier piling up of material within 1.5 AU due to the sweeping secular resonance when the gas disc's $\tau_{decay}$ = 1 Myr. Same as the low resolution case, the high resolution case ($r$ = 350 km) also forms Mars region embryos faster when $\tau_{decay}$ = 2 Myr, with 2 embryos growing beyond 1 $M_{\mathrm{mars}}$ within 10 Myr due to several giant impacts occurring after 5 Myr. It has to be mentioned that we have fewer simulations (only 2, both are EJS) for the high resolution classical model with $\tau_{decay}$ = 1 Myr than the high resolution simulations with $\tau_{decay}$ = 2 Myr (in total 4, combining EJS and CJS). In figure 13, we also present results which do not include a gas disc within the simulations (right panels). We could not find any of the Mars region embryos reaching 1 $M_{\mathrm{Mars}}$ in 10 Myr, suggesting the gas disc should likely exist at the beginning of planetesimal-planetesimal collision phase in order to form Mars fast.

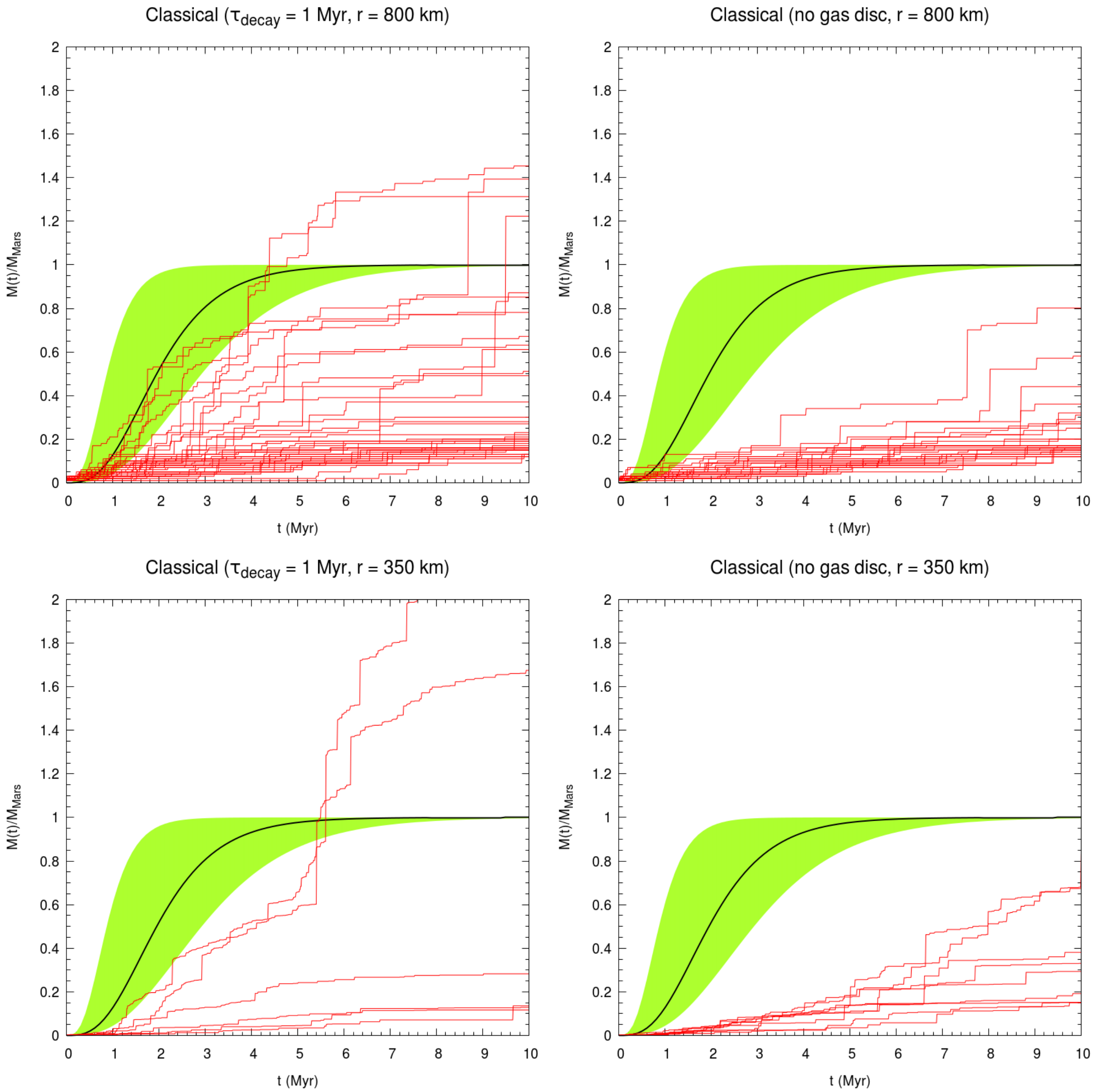

Figure 13 - Same as figure 9, but for $\tau_{decay}$ = 1 Myr (left panels) or without gas disc (right panels). Note the different range of the y-axis with figure 9.

## 4.2. Effect on embryos' accretion zones

### *4.2.1. EJS $\tau_{decay}$ = 1 Myr vs EJS $\tau_{decay}$ = 2 Myr*

Assuming a shorter lifetime of the gas disc, or no gas disc during embryo formation also affects the composition of the embryos. Figure 14 shows the accretion zone ($a_{\mathrm{weight}} \pm \sigma_{\mathrm{weight}}$, equation 15 and Equation 16) of the embryos in the Mercury, Venus, Earth and Mars regions for the EJS simulations with $\tau_{decay}$ = 2 Myr (upper left panel), EJS with $\tau_{decay}$ = 1 Myr (lower left panel), CJS with $\tau_{decay}$ = 2 Myr (upper right panel) and EJS without the gas disc (lower right panel). Comparing the two panels on the left, both cases form embryos with highly overlapping accretion zones. However, the Mercury region embryos formed in the EJS simulations with $\tau_{decay}$ = 1 Myr on average have lower $a_{\mathrm{weight}}$ than those in the EJS simulations with $\tau_{decay}$ = 2 Myr. This is supported by the mass percentage of material originating from different regions, which is shown in table 2 and table 5. In the classical simulations, embryos in the Mercury region in total accreted ~60% of their mass from < 1 AU when $\tau_{decay}$ = 1 Myr, compared to only ~25 % when $\tau_{decay}$ = 2 Myr. This is because most of the embryos that mainly accreted material < 1 AU survive to the end simulation when $\tau_{decay}$ = 1 Myr, whereas embryos formed early in the inner edge are lost to the Sun because of Type-I migration. The same explanation also applies to Venus, with embryos in the Venus region accreting twice as much material in terms of mass percentage from < 1 AU when $\tau_{decay}$ = 1 Myr than when $\tau_{decay}$ = 2 Myr. The mass percentage of materials accreted from > 2 AU is not greatly affected by the lifetime of the gas disc, with the difference in mass percentage less than 10% between $\tau_{decay}$ = 1 Myr and $\tau_{decay}$ = 2 Myr. This is because materials initially > 2 AU are implemented into the terrestrial planets region in the same way, regardless of the lifetime of the gas disc.

### *4.2.2. EJS vs CJS*

The orbits of the giant planets during the gas disc dissipation, on the other hand, have a larger impact on the implementation of material > 2 AU into the inner Solar system. With more circular orbits (CJS) assumed for Jupiter and Saturn, the mixing within the terrestrial region is much weaker than in the EJS case and hence the accretion zones of the embryos in CJS is much narrower than in EJS. The upper right panel of figure 14 depicts the accretion zones of embryos resulting in CJS simulation with $\tau_{decay}$ = 2 Myr. Compared to the corresponding EJS simulation (upper left of figure 14), the embryos in the CJS follow an obvious linear trend with a positive slope, whereas the data in the EJS simulations does not follow such a trend. This suggests that embryos accreted more locally in the CJS case than in the EJS case. This is explained by the weaker sweeping secular resonance effect when the gas giants are in more circular orbits (e.g. Morishima et al., 2010; Raymond et al., 2009). Figure 15 depicts the eccentricities versus semi-major axes of the objects in the CJS simulations (upper panels). Compared to the EJS case in figure 5, the whole disc is dynamically much colder ($e$ < 0.2 for most objects) during the gas disc dissipation and the sweeping secular resonance does not clear the asteroid belt region by pumping the eccentricities of the asteroid belt's object up. The implementation of asteroid belt's material into the inner Solar system (figure 6) is negligible in the CJS simulation. Hence, material > 2 AU is more difficult to be implemented into the terrestrial planet region in the CJS simulations. Both the Mercury and Venus region embryos do not have accretion zones extending to 2 AU in the CJS case.The $a_{\text{weight}}$ of the embryos in the Mars region is lying on the local accretion line $y = x$ in the CJS simulations (dashed line in figure 14). Closer to the Sun we found that the data points of the Mercury, Venus and Earth embryos are located above the dashed line in the CJS simulations. This is a result of the inward Type I migration of the embryos. We expect that the deviation between the $a_{\text{weight}}$ of embryos and the dashed line will be smaller if the gas disc's $\tau_{decay}$ is decreased to 1 Myr due to less effective inward migration.

#### *4.2.3. EJS with gas disc vs EJS without gas disc*

The existence of the gas disc is crucial to the composition of the embryos. The lower right panel of figure 14 shows the accretion zones of embryos from the EJS simulations, but without the gas disc. All embryos' $a_{\text{weight}}$ including the error bars lie on the linear correlation dashed line, suggesting an even stronger local accretion trend than in the CJS simulation with a gas disc. Regardless of the orbits of the giant planets, sweeping secular resonances cannot occur if the gas disc doesn't exist. Figure 15 depicts the eccentricities versus semi-major axes of the objects in the EJS simulations without the gas disc (lower panels). The disc is heated up much earlier when compared to the case with a dissipating gas disc in figure 5 because of the lack of gas drag. The excitement of eccentricities by the mean-motion resonances with Jupiter at ~2 AU (4:1), 2.5 AU (3:1) and 2.5 AU (5:2) is visible at 3 Myr. The location of this mean-motion resonances, however, does not change with time since Jupiter does not migrate in our simulation. Except for the locations in mean-motion resonances with Jupiter, most of the asteroid belt objects still exist beyond 2 AU. Therefore, mixing within a solid disc without a dissipating gas disc is relatively small, leading to obvious compositional differences between embryos in the innermost (Mercury) and the outermost (Mars) terrestrial planets' region. Table 5 shows a much higher contribution of material from < 1 AU to Mercury and Venus region embryos when a gas disc does not exist than any other cases. However, the clear trend developed for the accretion zones of the embryos without a gas disc usually does not persist on long timescales during the giant impact phase (Woo et al., 2018), so that the results of the accretion zones should be interpreted with caution.

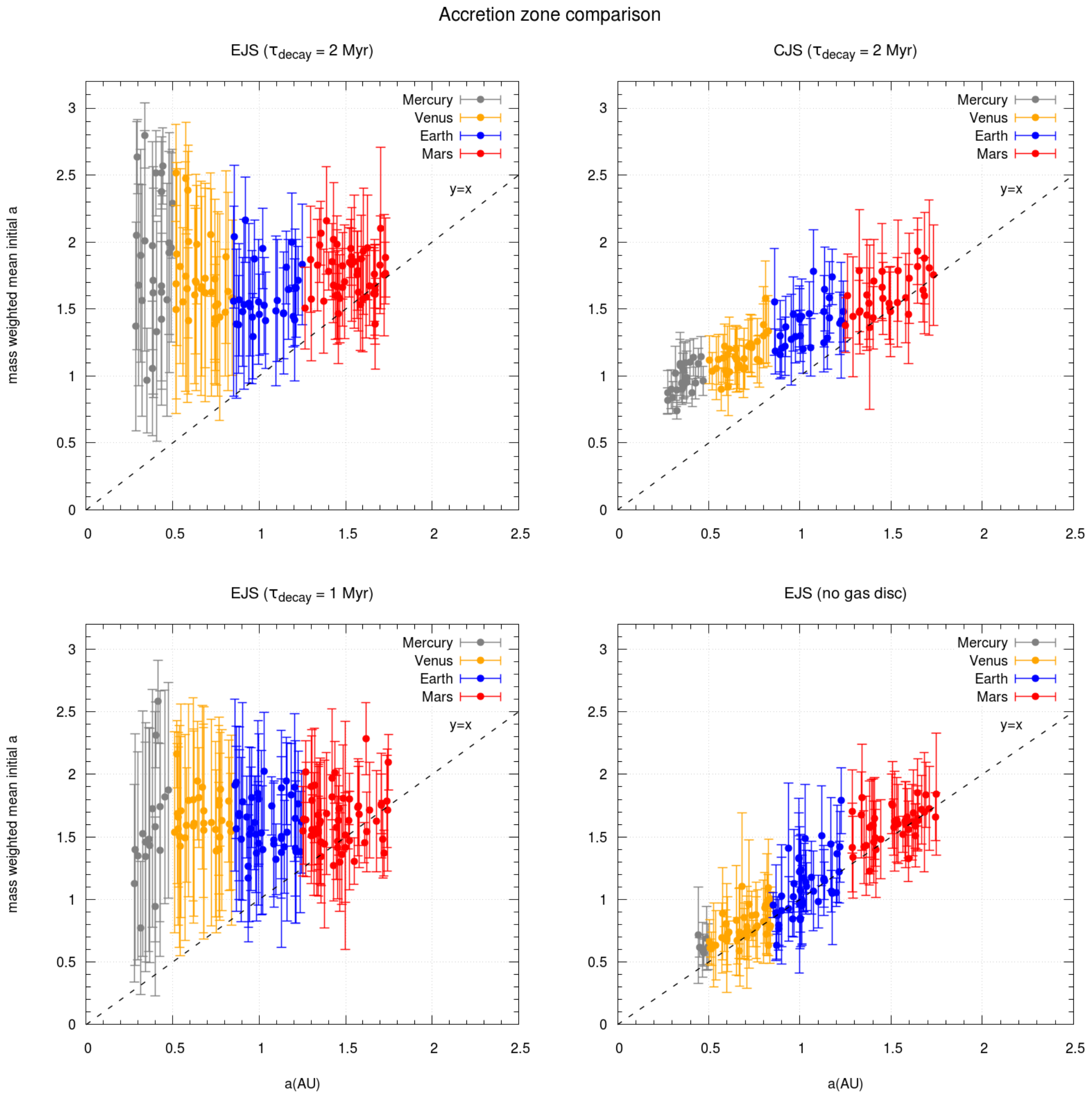

Figure 14 - Same as figure 7, but for comparison between results from simulation with initial eccentric Jupiter and Saturn (EJS) with gas disc's $\tau_{decay}$ = 2 Myr (top left panel), circular Jupiter and Saturn (CJS) with gas disc's $\tau_{decay}$ = 2 Myr (top right panel), EJS with gas disc's $\tau_{decay}$ = 1 Myr (lower left panel), and EJS without the gas disc (lower right panel).

Table 5 - Same as table 2, but for the EJS simulations with $\tau_{decay}$ = 1 Myr, CJS simulation with $\tau_{decay}$ = 2 Myr and EJS simulations without gas disc.

| | EJS ($\tau_{decay}$ = 1 Myr) | | |
|---|---|---|---|
| | a < 1 AU | 1 AU < a < 2 AU | a > 2 AU |
| Mercury | 59.9 % | 3.3 % | 37.0 % |
| Venus | 33.2 % | 30.0 % | 36.9 % |
| Earth | 17.6 % | 60.8 % | 21.7 % |
| Mars | 6.1 % | 79.4 % | 14.6 % |
| | CJS ($\tau_{decay}$ = 2 Myr) | | |
| | a < 1 AU | 1 AU < a < 2 AU | a > 2 AU |
| Mercury | 55.4 % | 44.6 % | 0 % |
| Venus | 23.2 % | 76.6 % | 0.3 % |
| Earth | 7.7 % | 89.7 % | 2.6 % |
| Mars | 3.2 % | 82.4 % | 14.5 % |
| | EJS (no gas disc) | | |
| | a < 1 AU | 1 AU < a < 2 AU | a > 2 AU |
| Mercury | 98.2 % | 0.5 % | 1.4 % |
| Venus | 92.5 % | 5.4 % | 2.1 % |
| Earth | 46.9 % | 48.3 % | 4.8 % |
| Mars | 0.7 % | 88.5 % | 11.1 % |

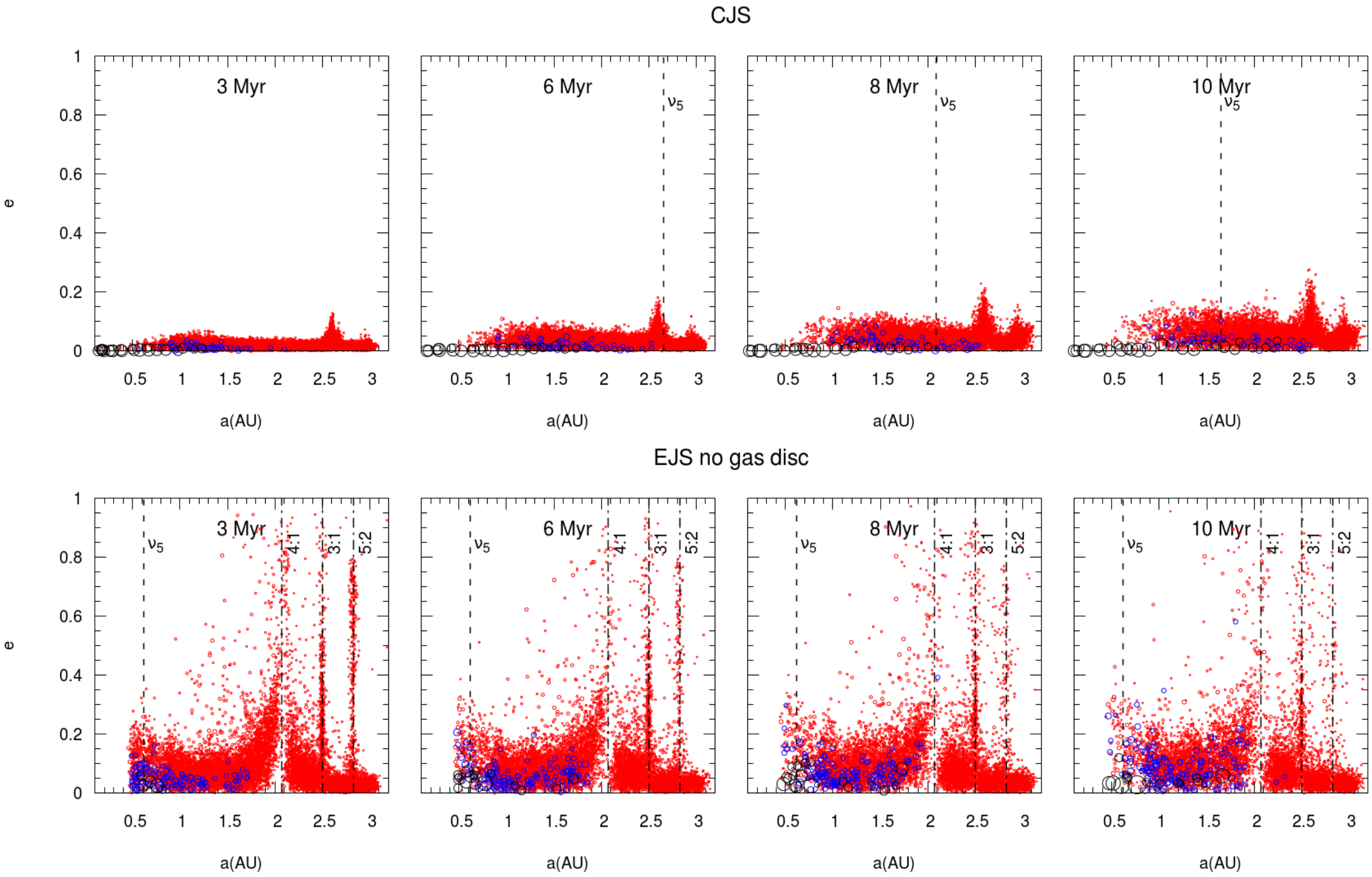

Figure 15 - Same as figure 5, but for the classical circular Jupiter and Saturn (CJS) model with $\tau_{decay}$ = 2 Myr (CL2) (upper panels) and initial eccentric Jupiter and Saturn (EJS) but without the gas disc (CL1_nogas). The location of the mean-motion resonances (4:1, 3:1 and 5:2) with Jupiter are also plotted in the lower panels.

## 5. Conclusions

In this paper, we report the outcomes of high-resolution N-body simulations of planetary embryo formation from a disc of equal mass planetesimals under the influence of the gas giants and protoplanetary disc for 10 Myr. All simulations were run with the GPU accelerated N-body code *GENGA* (Grimm and Stadel, 2014). We tested two models: the classical model, which assumes a solid surface density profile equal to the MMSN and the depleted disc model wherein the solid surface density is depleted relative to the MMSN beyond the current orbit of Mars. Compared to previous studies of embryo formation (e.g. Clement et al., 2020; Walsh and Levison, 2019), we focus more on comparing the embryo growth timescale and their composition with meteoritic data. Also, for most of our simulations we assumed initially eccentric Jupiter and Saturn (EJS), while the previous studies assumed more circular orbits for the gas giants (CJS).

### 5.1. Strong inside-out growth reproduced

When the gas dissipation timescale is 2 Myr we produce ~20 lunar- to Mars-mass embryos in the current Mercury, Venus and Earth regions. The mass of embryos in the Mars region are almost always slightly smaller in the depleted disc model than in the classical model after 10 Myr of growth. The time evolution of the percentage distribution of mass in embryos, proto-embryos and planetesimals are similar in the classical and depleted disc models, with embryo masses reaching ~80 % of the disc mass within 2 Myr in the region < 1 AU. Planetesimal masses, on the other hand, accounting for 80% to 100 % of the disc > 2 AU. This suggests that our simulations reproduce the strong inside-out growth of embryos in previous studies (e.g Carter et al. 2015; Clement et al. 2020; Morishima et al. 2010; Walsh et al. 2019) even though we apply different initial conditions. A bimodal mass distribution has not been achieved in the intermediate region of the disc (1 AU < *a* < 2 AU) within 10 Myr, with more than 20% of the total mass residing in intermediate masses objects - proto-embryos ranging from 0.001 to 0.01 $M_{Earth}$ - in both models.

### 5.2. Embryo growth timescale is affected by initial planetesimal size

The time for embryos to reach a lunar mass are similar in the classical and depleted disc models. Assuming different initial radii of the planetesimals (either *r* = 350 km or 800 km), however, results in a different formation time scale. In general, embryos ending in the Mercury and Venus regions reach a lunar mass faster in high resolution simulations (initial *r* = 350 km) than low resolution runs (initial *r* = 800 km). It takes on average 2 to 3 Myr in the high resolution simulations but can last longer than 8 Myr in the low resolution cases. This is explained by the loss of the early-forming embryos due to Type-I migration and slow-forming embryos originating from the asteroid belt region are implanted into the inner Solar system by the sweeping secular resonance (Bromley & Keyon 2017**;** Heppenheimer, 1980; Lecar and Franklin, 1997; Nagasawa et al., 2001, 2000; Ward, 1981) in the low resolution case. The time reaching lunar mass for embryos in the Earth and Mars region has less difference between the low and high resolution cases. The mean time for Earth's region embryos reaching a lunar mass is 3 to 4 Myr in both models. On average Mars region embryos take 1 to 3 Myr longer to form than Earth's region embryos. Our results are broadly consistent with previous studies (Clement et al. 2020;Walsh et al. 2019)**.** Our results also indicate a slower embryo formation rate than that proposed for proto-Venus and proto-Earth based on atmospheric noble gas isotopic ratios, volatile abundances (Lammer et al., 2020) and pebble accretion (Schiller et al., 2018, 2020), although these proposed constraints are still under debate.

### 5.3. Criteria for fast growth of Mars

Compared to Venus and Earth, Mars' formation timescale has a stricter constraint, with the Hf-W and Fe-Ni isotopic chronology implying a rapid formation of Mars, reaching half of its mass at about 2 Myr after the birth of the Solar system; the planet was fully formed within 10 Myr (Dauphas and Pourmand, 2011; Tang and Dauphas, 2014; cf. Marchi et al., 2020). We thus make a comparison between our N-body simulation results and the proposed formation time scale for Mars from Hf-W chronology of Dauphas and Pourmand (2011). We found that all the embryos in the Mars region grow slower than the proposed rate from Hf-W chronology in both the classical and depleted disc models with $\tau_{decay}$ = 2 Myr. The fastest-growing embryo reaches 0.5 $M_{Mars}$ after ~4 Myr and 1 $M_{Mars}$ after ~7 Myr. Based on the results of our simulations, there are two ways to speed up the growth of Mars: Mars mainly formed from larger-sized planetesimals (i.e. low resolution simulations) with massive impactor (> 0.1 $M_{target}$) contributing 40 to 50 % of the total mass, and we assume a shorter gas disc dissipation time (e.g. $\tau_{decay}$ = 1 Myr) with Jupiter and Saturn on EJS orbits. The shorter $\tau_{decay}$ causes the earlier occurrence of the sweeping secular resonance, which results in a piling up of material in the terrestrial planet region several Myr earlier than when $\tau_{decay}$ = 2 Myr. When assuming an initial planetesimal radius $r$ = 800 km and $\tau_{decay}$ = 1 Myr for the gas disc, the EJS classical model simulation results in Mars region embryos reaching 0.5 $M_{\mathrm{Mars}}$ after ~2 Myr and 1 $M_{\mathrm{Mars}}$ before 5 Myr, which matches the Hf-W chronology reasonably well. In theory, the formation of embryos can also be sped up by implementing fragmentation into the N-body code and lowering the diameter of the initial planetesimals so that the fragment gravitational cross section of the embryos are increased (Kobayashi and Dauphas, 2013; Walsh and Levison, 2019), a higher solid surface density profile of the disc (Kobayashi and Dauphas, 2013) and/or inducing a slight excitement of the dynamical state to the planetesimals by gas disc turbulence (Laughlin et al., 2004; Ogihara et al., 2007).

### 5.4. Sweeping secular resonance alters the composition of embryos

The sweeping secular resonance also affects the composition of the embryos. We found that material originating beyond 2 AU is constantly implemented into the terrestrial planet region during the dissipation of the gas disc. Most of the material is implemented into the Mercury and Venus region, with more than 35% of materials in the embryos of Mercury and Venus originating beyond 2 AU. The sweeping secular resonance together with the strong Type-I migration effect enhanced the mixing within the terrestrial planet regions, resulting in similar accretion (feeding) zones of the embryos formed all over the terrestrial planet region in both the EJS classical and depleted disc models. Lowering the gas dissipation timescale to 1 Myr results in similar highly overlapping accretion zones for the embryos, but increasing the percentage of mass originated from < 1 AU for Mercury and Venus for more than twice. On the other hand, assuming circular orbits for the giant planets or without including a gas disc in the simulation decreases the mixing effect greatly because of the negligible effect of the sweeping secular resonance. As such, the accretion zones of the embryos formed in the terrestrial regions end in a positive linear trend (i.e. inner embryos accreted mainly material closer to the Sun and the outer embryos mainly accreted material further away). If a spatial isotopic composition gradient existed initially right before the formation of planetesimals (e.g. Ek et al., 2020; Fischer-Gödde and Kleine, 2017; Render et al., 2017; Yamakawa et al., 2010), then we expect such a gradient to be preserved after the formation of embryos when the orbits of Jupiter and Saturn are more circular than their current orbits during gas disc dissipation or the embryos form in a gas free condition.

### 5.5. Samples from Mercury and Venus can constrain initial gas giant architecture

Our N-body simulation results suggested that a shorter Solar nebula decay timescale $\tau_{decay}$ (≤ 1 Myr), which matches the meteoritic magnetism data (Wang et al., 2017) and the Pb-Pb dating of chondrules (Bollard et al., 2017), is more likely to explain the fast formation of Mars if the gas giants possessed their current orbits (EJS) during gas disc's dissipation. The bulk composition of embryos formed in Mercury and Venus region are highly dependent on the gas disc properties and the giant planets' orbits during gas disc dissipation. Although longer simulations are required to compute the final bulk composition of the terrestrial planets, based on the current results we propose that samples from Mercury and Venus could aid us in deducing the condition of the gas disc during planetesimal and embryo formation, the lifetime of the gas disc and the orbital state of the giant planets when the gas disc dissipated.

## 6. Acknowledgements

This work has been carried out within the framework of the National Center of Competence in Research PlanetS, supported by the Swiss National Science Foundation (SNSF). The authors acknowledge the financial support of the SNSF. The authors acknowledge the computational support from Service and Support for Science IT ($S^3IT$) of University of Zurich and Swiss National Supercomputing Centre (CSCS). RB acknowledges financial assistance from the Japan Society for the Promotion of Science (JSPS) Shingakujutsu Kobo (JP19H05071). JMYW thanks Nicolas Dauphas and Man Hoi Lee for discussions and Volker Hoffmann for technical assistance.

## References

Agnor, C.B., Canup, R.M., Levison, H.F., 1999. On the Character and Consequences of Large Impacts in the Late Stage of Terrestrial Planet Formation. Icarus 142, 219–237. https://doi.org/10.1006/icar.1999.6201

Armitage, P.J., Clarke, C.J., Palla, F., 2003. Dispersion in the lifetime and accretion rate of T Tauri discs. Mon. Not. R. Astron. Soc. 342, 1139–1146. https://doi.org/10.1046/j.1365-8711.2003.06604.x

Benz, W., 2000. Low Velocity Collisions and the Growth of Planetesimals, in: Benz, W., Kallenbach, R., Lugmair, G.W. (Eds.), From Dust to Terrestrial Planets, Space Sciences Series of ISSI. Springer Netherlands, Dordrecht, pp. 279–294. https://doi.org/10.1007/978-94-011-4146-8_18

Benz, W., Asphaug, E., 1999. Catastrophic Disruptions Revisited. Icarus 142, 5–20. https://doi.org/10.1006/icar.1999.6204

Bollard, J., Connelly, J.N., Whitehouse, M.J., Pringle, E.A., Bonal, L., Jørgensen, J.K., Nordlund, Å., Moynier, F., Bizzarro, M., 2017. Early formation of planetary building blocks inferred from Pb isotopic ages of chondrules. Sci. Adv. 3, e1700407. https://doi.org/10.1126/sciadv.1700407

Borg, L.E., Brennecka, G.A., Symes, S.J.K., 2016. Accretion timescale and impact history of Mars deduced from the isotopic systematics of martian meteorites. Geochim. Cosmochim. Acta 175, 150–167. https://doi.org/10.1016/j.gca.2015.12.002

Boss, A.P., 1997. Giant Planet Formation by Gravitational Instability. Science 276, 1836–1839. https://doi.org/10.1126/science.276.5320.1836

Bouvier, A., Wadhwa, M., 2010. The age of the Solar System redefined by the oldest Pb–Pb age of a meteoritic inclusion. Nat. Geosci. 3, 637–641. https://doi.org/10.1038/ngeo941

Bouvier, L.C., Costa, M.M., Connelly, J.N., Jensen, N.K., Wielandt, D., Storey, M., Nemchin, A.A., Whitehouse, M.J., Snape, J.F., Bellucci, J.J., Moynier, F., Agranier, A., Gueguen, B., Schönbächler, M., Bizzarro, M., 2018. Evidence for extremely rapid magma ocean crystallization and crust formation on Mars. Nature 558, 586–589. https://doi.org/10.1038/s41586-018-0222-z

Brasser, R., Dauphas, N., Mojzsis, S.J., 2018. Jupiter's Influence on the Building Blocks of Mars and Earth. Geophys. Res. Lett. 45, 5908–5917. https://doi.org/10.1029/2018GL078011

Brasser, R., Matsumura, S., Ida, S., Mojzsis, S.J., Werner, S.C., 2016. ANALYSIS OF TERRESTRIAL PLANET FORMATION BY THE GRAND TACK MODEL: SYSTEM ARCHITECTURE AND TACK LOCATION. Astrophys. J. 821, 75. https://doi.org/10.3847/0004-637X/821/2/75

Brasser, R., Mojzsis, S.J., 2020. The partitioning of the inner and outer Solar System by a structured protoplanetary disk. Nat. Astron. 4, 492–499. https://doi.org/10.1038/s41550-019-0978-6

Brasser, R., Mojzsis, S.J., Matsumura, S., Ida, S., 2017. The cool and distant formation of Mars. Earth Planet. Sci. Lett. 468, 85–93. https://doi.org/10.1016/j.epsl.2017.04.005
Bromley, B.C., Kenyon, S.J., 2017. Terrestrial Planet Formation: Dynamical Shake-up and the Low Mass of Mars. Astron. J. 153, 216. https://doi.org/10.3847/1538-3881/aa6aaa
Canup, R.M., Asphaug, E., 2001. Origin of the Moon in a giant impact near the end of the Earth's formation. Nature 412, 708–712. https://doi.org/10.1038/35089010
Carlson, R.W., Brasser, R., Yin, Q.-Z., Fischer-Gödde, M., Qin, L., 2018. Feedstocks of the Terrestrial Planets. Space Sci. Rev. 214, 121. https://doi.org/10.1007/s11214-018-0554-x
Carter, P.J., Leinhardt, Z.M., Elliott, T., Walter, M.J., Stewart, S.T., 2015. COMPOSITIONAL EVOLUTION DURING ROCKY PROTOPLANET ACCRETION. Astrophys. J. 813, 72. https://doi.org/10.1088/0004-637X/813/1/72
Chambers, J., 2006. A semi-analytic model for oligarchic growth. Icarus 180, 496–513. https://doi.org/10.1016/j.icarus.2005.10.017
Chambers, J.E., 2001. Making More Terrestrial Planets. Icarus 152, 205–224. https://doi.org/10.1006/icar.2001.6639
Chambers, J.E., 1999. A hybrid symplectic integrator that permits close encounters between massive bodies. Mon. Not. R. Astron. Soc. 304, 793–799. https://doi.org/10.1046/j.1365-8711.1999.02379.x
Clement, M.S., Kaib, N.A., Chambers, J.E., 2020. Embryo Formation with GPU Acceleration: Reevaluating the Initial Conditions for Terrestrial Accretion. Planet. Sci. J. 1, 18. https://doi.org/10.3847/PSJ/ab91aa
Clement, M.S., Kaib, N.A., Raymond, S.N., Chambers, J.E., Walsh, K.J., 2019. The early instability scenario: Terrestrial planet formation during the giant planet instability, and the effect of collisional fragmentation. Icarus 321, 778–790. https://doi.org/10.1016/j.icarus.2018.12.033
Clement, M.S., Kaib, N.A., Raymond, S.N., Walsh, K.J., 2018. Mars' growth stunted by an early giant planet instability. Icarus 311, 340–356. https://doi.org/10.1016/j.icarus.2018.04.008
Costa, M.M., Jensen, N.K., Bouvier, L.C., Connelly, J.N., Mikouchi, T., Horstwood, M.S.A., Suuronen, J.-P., Moynier, F., Deng, Z., Agranier, A., Martin, L.A.J., Johnson, T.E., Nemchin, A.A., Bizzarro, M., 2020. The internal structure and geodynamics of Mars inferred from a 4.2-Gyr zircon record. Proc. Natl. Acad. Sci. 117, 30973–30979. https://doi.org/10.1073/pnas.2016326117
Cresswell, P., Dirksen, G., Kley, W., Nelson, R.P., 2007. On the evolution of eccentric and inclined protoplanets embedded in protoplanetary disks. Astron. Astrophys. 473, 329–342. https://doi.org/10.1051/0004-6361:20077666
Cuzzi, J.N., Hogan, R.C., Shariff, K., 2008. Toward Planetesimals: Dense Chondrule Clumps in the Protoplanetary Nebula. Astrophys. J. 687, 1432–1447. https://doi.org/10.1086/591239
Dauphas, N., 2017. The isotopic nature of the Earth's accreting material through time. Nature 541, 521–524. https://doi.org/10.1038/nature20830
Dauphas, N., Chen, J.H., Zhang, J., Papanastassiou, D.A., Davis, A.M., Travaglio, C., 2014. Calcium-48 isotopic anomalies in bulk chondrites and achondrites: Evidence for a uniform isotopic reservoir in the inner protoplanetary disk. Earth Planet. Sci. Lett. 407, 96–108. https://doi.org/10.1016/j.epsl.2014.09.015
Dauphas, N., Davis, A.M., Marty, B., Reisberg, L., 2004. The cosmic molybdenum–ruthenium isotope correlation. Earth Planet. Sci. Lett. 226, 465–475. https://doi.org/10.1016/j.epsl.2004.07.026
Dauphas, N., Pourmand, A., 2011. Hf–W–Th evidence for rapid growth of Mars and its status as a planetary embryo. Nature 473, 489–492. https://doi.org/10.1038/nature10077
Davis, A.M., 2005. Meteorites, Comets, and Planets: Treatise on Geochemistry, Second Edition, Volume 1. Elsevier.
Debaille, V., Brandon, A.D., Yin, Q.Z., Jacobsen, B., 2007. Coupled 142 Nd– 143 Nd evidence for a protracted magma ocean in Mars. Nature 450, 525–528. https://doi.org/10.1038/nature06317

Deguen, R., Olson, P., Cardin, P., 2011. Experiments on turbulent metal-silicate mixing in a magma ocean. Earth Planet. Sci. Lett. 310, 303–313. https://doi.org/10.1016/j.epsl.2011.08.041
Delbo, M., Avdellidou, C., Morbidelli, A., 2019. Ancient and primordial collisional families as the main sources of X-type asteroids of the inner main belt. Astron. Astrophys. 624, A69. https://doi.org/10.1051/0004-6361/201834745
Delbo, M., Walsh, K., Bolin, B., Avdellidou, C., Morbidelli, A., 2017. Identification of a primordial asteroid family constrains the original planetesimal population. Science 357, 1026–1029. https://doi.org/10.1126/science.aam6036
Dominik, C., Blum, J., Cuzzi, J.N., Wurm, G., 2007. Growth of Dust as the Initial Step Toward Planet Formation. Protostars Planets V 783–800.
Donahue, T.M., 1986. Fractionation of noble gases by thermal escape from accreting planetesimals. Icarus 66, 195–210. https://doi.org/10.1016/0019-1035(86)90151-X
Drążkowska, J., Alibert, Y., Moore, B., 2016. Close-in planetesimal formation by pile-up of drifting pebbles. Astron. Astrophys. 594, A105. https://doi.org/10.1051/0004-6361/201628983
Duncan, M.J., Levison, H.F., Lee, M.H., 1998. A Multiple Time Step Symplectic Algorithm for Integrating Close Encounters. Astron. J. 116, 2067. https://doi.org/10.1086/300541
Durisen, R.H., Boss, A.P., Mayer, L., Nelson, A.F., Quinn, T., Rice, W.K.M., 2007. Gravitational Instabilities in Gaseous Protoplanetary Disks and Implications for Giant Planet Formation. Protostars Planets V 607–622.
Ek, M., Hunt, A.C., Lugaro, M., Schönbächler, M., 2020. The origin of s -process isotope heterogeneity in the solar protoplanetary disk. Nat. Astron. 4, 273–281. https://doi.org/10.1038/s41550-019-0948-z
Fischer, R.A., Nimmo, F., O'Brien, D.P., 2018. Radial mixing and Ru–Mo isotope systematics under different accretion scenarios. Earth Planet. Sci. Lett. 482, 105–114. https://doi.org/10.1016/j.epsl.2017.10.055
Fischer-Gödde, M., Kleine, T., 2017. Ruthenium isotopic evidence for an inner Solar System origin of the late veneer. Nature 541, 525–527. https://doi.org/10.1038/nature21045
Fitoussi, C., Bourdon, B., Wang, X., 2016. The building blocks of Earth and Mars: A close genetic link. Earth Planet. Sci. Lett. 434, 151–160. https://doi.org/10.1016/j.epsl.2015.11.036
Foley, C.N., Wadhwa, M., Borg, L.E., Janney, P.E., Hines, R., Grove, T.L., 2005. The early differentiation history of Mars from 182W-142Nd isotope systematics in the SNC meteorites. Geochim. Cosmochim. Acta 69, 4557–4571. https://doi.org/10.1016/j.gca.2005.05.009
Grimm, S.L., Stadel, J.G., 2014. THE GENGA CODE: GRAVITATIONAL ENCOUNTERS INN-BODY SIMULATIONS WITH GPU ACCELERATION. Astrophys. J. 796, 23. https://doi.org/10.1088/0004-637X/796/1/23
Hansen, B.M.S., 2009. FORMATION OF THE TERRESTRIAL PLANETS FROM A NARROW ANNULUS. Astrophys. J. 703, 1131–1140. https://doi.org/10.1088/0004-637X/703/1/1131
Harrison, T.M., Schmitt, A.K., McCulloch, M.T., Lovera, O.M., 2008. Early (≥4.5 Ga) formation of terrestrial crust: Lu–Hf, δ18O, and Ti thermometry results for Hadean zircons. Earth Planet. Sci. Lett. 268, 476–486. https://doi.org/10.1016/j.epsl.2008.02.011
Hayashi, C., 1981. Structure of the Solar Nebula, Growth and Decay of Magnetic Fields and Effects of Magnetic and Turbulent Viscosities on the Nebula. Prog. Theor. Phys. Suppl. 70, 35–53. https://doi.org/10.1143/PTPS.70.35
Heppenheimer, T.A., 1980. Secular resonances and the origin of eccentricities of Mars and the asteroids. Icarus 41, 76–88. https://doi.org/10.1016/0019-1035(80)90160-8
Hoffman, J.H., Hodges, R.R., Donahue, T.M., McElroy, M.B., 1980. Composition of the Venus lower atmosphere from the Pioneer Venus Mass Spectrometer. J. Geophys. Res. Space Phys. 85, 7882–7890. https://doi.org/10.1029/JA085iA13p07882
Hoffmann, V., Grimm, S.L., Moore, B., Stadel, J., 2017. Stochasticity and predictability in terrestrial planet formation. Mon. Not. R. Astron. Soc. 465, 2170–2188. https://doi.org/10.1093/mnras/stw2856

Hunten, D.M., Pepin, R.O., Walker, J.C.G., 1987. Mass fractionation in hydrodynamic escape. Icarus 69, 532–549. https://doi.org/10.1016/0019-1035(87)90022-4

Ida, S., Makino, J., 1993. Scattering of Planetesimals by a Protoplanet: Slowing Down of Runaway Growth. Icarus 106, 210–227. https://doi.org/10.1006/icar.1993.1167

Ikoma, M., Genda, H., 2006. Constraints on the Mass of a Habitable Planet with Water of Nebular Origin. Astrophys. J. 648, 696. https://doi.org/10.1086/505780

Istomin, V.G., Grechnev, K.V., Kochnev, V.A., 1983. Venera 13 and 14 - Mass spectroscopy of the atmosphere. Kosmicheskie Issled. 21, 410–420.

Istomin, V.G., Grechnev, K.V., Kotchnev, V.A., 1980. Mass Spectrometer Measurements of the Composition of the Lower Atmosphere of Venus, in: Rycroft, M.J. (Ed.), COSPAR Colloquia Series, Space Research. Pergamon, pp. 215–218. https://doi.org/10.1016/S0964-2749(13)60044-X

Izidoro, A., Haghighipour, N., Winter, O.C., Tsuchida, M., 2014. TERRESTRIAL PLANET FORMATION IN A PROTOPLANETARY DISK WITH A LOCAL MASS DEPLETION: A SUCCESSFUL SCENARIO FOR THE FORMATION OF MARS. Astrophys. J. 782, 31. https://doi.org/10.1088/0004-637X/782/1/31

Izidoro, A., Raymond, S.N., Morbidelli, A., Winter, O.C., 2015. Terrestrial planet formation constrained by Mars and the structure of the asteroid belt. Mon. Not. R. Astron. Soc. 453, 3619–3634. https://doi.org/10.1093/mnras/stv1835

Jacobson, S.A., Morbidelli, A., 2014. Lunar and terrestrial planet formation in the Grand Tack scenario. Philos. Trans. R. Soc. Math. Phys. Eng. Sci. 372, 20130174. https://doi.org/10.1098/rsta.2013.0174

Jin, L., Arnett, W.D., Sui, N., Wang, X., 2008. An Interpretation of the Anomalously Low Mass of Mars. Astrophys. J. Lett. 674, L105. https://doi.org/10.1086/529375

Johansen, A., Low, M.-M.M., Lacerda, P., Bizzarro, M., 2015. Growth of asteroids, planetary embryos, and Kuiper belt objects by chondrule accretion. Sci. Adv. 1, e1500109. https://doi.org/10.1126/sciadv.1500109

Johansen, A., Oishi, J.S., Low, M.-M.M., Klahr, H., Henning, T., Youdin, A., 2007. Rapid planetesimal formation in turbulent circumstellar disks. Nature 448, 1022–1025. https://doi.org/10.1038/nature06086

Johnstone, C.P., Güdel, M., Stökl, A., Lammer, H., Tu, L., Kislyakova, K.G., Lüftinger, T., Odert, P., Erkaev, N.V., Dorfi, E.A., 2015. THE EVOLUTION OF STELLAR ROTATION AND THE HYDROGEN ATMOSPHERES OF HABITABLE-ZONE TERRESTRIAL PLANETS. Astrophys. J. 815, L12. https://doi.org/10.1088/2041-8205/815/1/L12

Kaib, N.A., Cowan, N.B., 2015. The feeding zones of terrestrial planets and insights into Moon formation. Icarus 252, 161–174. https://doi.org/10.1016/j.icarus.2015.01.013

Karl E. Haisch, Jr., Lada, E.A., Lada, C.J., 2001. Disk Frequencies and Lifetimes in Young Clusters. Astrophys. J. 553, L153–L156. https://doi.org/10.1086/320685

Kenyon, S.J., Bromley, B.C., 2006. Terrestrial Planet Formation. I. The Transition from Oligarchic Growth to Chaotic Growth*. Astron. J. 131, 1837. https://doi.org/10.1086/499807

Kleine, T., Münker, C., Mezger, K., Palme, H., 2002. Rapid accretion and early core formation on asteroids and the terrestrial planets from Hf–W chronometry. Nature 418, 952–955. https://doi.org/10.1038/nature00982

Kleine, T., Touboul, M., Bourdon, B., Nimmo, F., Mezger, K., Palme, H., Jacobsen, S.B., Yin, Q.-Z., Halliday, A.N., 2009. Hf–W chronology of the accretion and early evolution of asteroids and terrestrial planets. Geochim. Cosmochim. Acta, The Chronology of Meteorites and the Early Solar System 73, 5150–5188. https://doi.org/10.1016/j.gca.2008.11.047

Kobayashi, H., Dauphas, N., 2013. Small planetesimals in a massive disk formed Mars. Icarus 225, 122–130. https://doi.org/10.1016/j.icarus.2013.03.006

Kobayashi, H., Tanaka, H., Krivov, A.V., Inaba, S., 2010. Planetary growth with collisional fragmentation and gas drag. Icarus 209, 836–847. https://doi.org/10.1016/j.icarus.2010.04.021

Kokubo, E., Ida, S., 2002. Formation of Protoplanet Systems and Diversity of Planetary Systems. Astrophys. J. 581, 666. https://doi.org/10.1086/344105
Kokubo, E., Ida, S., 2000. Formation of Protoplanets from Planetesimals in the Solar Nebula. Icarus 143, 15–27. https://doi.org/10.1006/icar.1999.6237
Kokubo, E., Ida, S., 1998. Oligarchic Growth of Protoplanets. Icarus 131, 171–178. https://doi.org/10.1006/icar.1997.5840
Kokubo, E., Ida, S., 1996. On Runaway Growth of Planetesimals. Icarus 123, 180–191. https://doi.org/10.1006/icar.1996.0148
Kokubo, E., Ida, S., 1995. Orbital Evolution of Protoplanets Embedded in a Swarm of Planetesimals. Icarus 114, 247–257. https://doi.org/10.1006/icar.1995.1059
Kruijer, T.S., Borg, L.E., Wimpenny, J., Sio, C.K., 2020. Onset of magma ocean solidification on Mars inferred from Mn-Cr chronometry. Earth Planet. Sci. Lett. 542, 116315. https://doi.org/10.1016/j.epsl.2020.116315
Kruijer, T.S., Burkhardt, C., Budde, G., Kleine, T., 2017a. Age of Jupiter inferred from the distinct genetics and formation times of meteorites. Proc. Natl. Acad. Sci. 114, 6712–6716. https://doi.org/10.1073/pnas.1704461114
Kruijer, T.S., Kleine, T., Borg, L.E., Brennecka, G.A., Irving, A.J., Bischoff, A., Agee, C.B., 2017b. The early differentiation of Mars inferred from Hf–W chronometry. Earth Planet. Sci. Lett. 474, 345–354. https://doi.org/10.1016/j.epsl.2017.06.047
Kruijer, T.S., Touboul, M., Fischer-Gödde, M., Bermingham, K.R., Walker, R.J., Kleine, T., 2014. Protracted core formation and rapid accretion of protoplanets. Science 344, 1150–1154. https://doi.org/10.1126/science.1251766
Lambrechts, M., Johansen, A., 2012. Rapid growth of gas-giant cores by pebble accretion. Astron. Astrophys. 544, A32. https://doi.org/10.1051/0004-6361/201219127
Lammer, H., Leitzinger, M., Scherf, M., Odert, P., Burger, C., Kubyshkina, D., Johnstone, C., Maindl, T., Schäfer, C.M., Güdel, M., Tosi, N., Nikolaou, A., Marcq, E., Erkaev, N.V., Noack, L., Kislyakova, K.G., Fossati, L., Pilat-Lohinger, E., Ragossnig, F., Dorfi, E.A., 2020. Constraining the early evolution of Venus and Earth through atmospheric Ar, Ne isotope and bulk K/U ratios. Icarus 339, 113551. https://doi.org/10.1016/j.icarus.2019.113551
Lammer, H., Stökl, A., Erkaev, N.V., Dorfi, E.A., Odert, P., Güdel, M., Kulikov, Y.N., Kislyakova, K.G., Leitzinger, M., 2014. Origin and loss of nebula-captured hydrogen envelopes from 'sub'- to 'super-Earths' in the habitable zone of Sun-like stars. Mon. Not. R. Astron. Soc. 439, 3225–3238. https://doi.org/10.1093/mnras/stu085
Laughlin, G., Steinacker, A., Adams, F.C., 2004. Type I Planetary Migration with MHD Turbulence. Astrophys. J. 608, 489–496. https://doi.org/10.1086/386316
Lecar, M., Franklin, F., 1997. The Solar Nebula, Secular Resonances, Gas Drag, and the Asteroid Belt. Icarus 129, 134–146. https://doi.org/10.1006/icar.1997.5782
Lee, D.-C., Halliday, A.N., 1997. Core formation on Mars and differentiated asteroids. Nature 388, 854–857. https://doi.org/10.1038/42206
Leinhardt, Z.M., Richardson, D.C., Lufkin, G., Haseltine, J., 2009. Planetesimals to protoplanets – II. Effect of debris on terrestrial planet formation. Mon. Not. R. Astron. Soc. 396, 718–728. https://doi.org/10.1111/j.1365-2966.2009.14769.x
Levison, H.F., Duncan, M.J., Thommes, E., 2012. A LAGRANGIAN INTEGRATOR FOR PLANETARY ACCRETION AND DYNAMICS (LIPAD). Astron. J. 144, 119. https://doi.org/10.1088/0004-6256/144/4/119
Levison, H.F., Kretke, K.A., Duncan, M.J., 2015a. Growing the gas-giant planets by the gradual accumulation of pebbles. Nature 524, 322–324. https://doi.org/10.1038/nature14675
Levison, H.F., Kretke, K.A., Walsh, K.J., Bottke, W.F., 2015b. Growing the terrestrial planets from the gradual accumulation of submeter-sized objects. Proc. Natl. Acad. Sci. 112, 14180–14185. https://doi.org/10.1073/pnas.1513364112

Lissauer, J.J., 1993. Planet Formation. Annu. Rev. Astron. Astrophys. 31, 129–172. https://doi.org/10.1146/annurev.aa.31.090193.001021
Lissauer, J.J., Hubickyj, O., D'Angelo, G., Bodenheimer, P., 2009. Models of Jupiter's growth incorporating thermal and hydrodynamic constraints. Icarus 199, 338–350. https://doi.org/10.1016/j.icarus.2008.10.004
Lodders, K., Fegley, B., 1997. An Oxygen Isotope Model for the Composition of Mars. Icarus 126, 373–394. https://doi.org/10.1006/icar.1996.5653
Lykawka, P.S., Ito, T., 2019. Constraining the Formation of the Four Terrestrial Planets in the Solar System. Astrophys. J. 883, 130. https://doi.org/10.3847/1538-4357/ab3b0a
Lykawka, P.S., Ito, T., 2013. TERRESTRIAL PLANET FORMATION DURING THE MIGRATION AND RESONANCE CROSSINGS OF THE GIANT PLANETS. Astrophys. J. 773, 65. https://doi.org/10.1088/0004-637X/773/1/65
Mah, J., Brasser, R., 2021. Isotopically distinct terrestrial planets via local accretion. Icarus 354, 114052. https://doi.org/10.1016/j.icarus.2020.114052
Marchi, S., Walker, R.J., Canup, R.M., 2020. A compositionally heterogeneous martian mantle due to late accretion. Sci. Adv. 6, eaay2338. https://doi.org/10.1126/sciadv.aay2338
Mezger, K., Debaille, V., Kleine, T., 2013. Core Formation and Mantle Differentiation on Mars. Space Sci. Rev. 174, 27–48. https://doi.org/10.1007/s11214-012-9935-8
Mezger, K., Schönbächler, M., Bouvier, A., 2020. Accretion of the Earth—Missing Components? Space Sci. Rev. 216, 27. https://doi.org/10.1007/s11214-020-00649-y
Mojzsis, S.J., Brasser, R., Kelly, N.M., Abramov, O., Werner, S.C., 2019. Onset of Giant Planet Migration before 4480 Million Years Ago. Astrophys. J. 881, 44. https://doi.org/10.3847/1538-4357/ab2c03
Morbidelli, A., 2018. Calcium signals in planetary embryos. Nature 555, 451–452. https://doi.org/10.1038/d41586-018-03144-1
Morbidelli, A., Bottke, W.F., Nesvorný, D., Levison, H.F., 2009. Asteroids were born big. Icarus 204, 558–573. https://doi.org/10.1016/j.icarus.2009.07.011
Morishima, R., 2015. A particle-based hybrid code for planet formation. Icarus 260, 368–395. https://doi.org/10.1016/j.icarus.2015.07.030
Morishima, R., Golabek, G.J., Samuel, H., 2013. N-body simulations of oligarchic growth of Mars: Implications for Hf–W chronology. Earth Planet. Sci. Lett. 366, 6–16. https://doi.org/10.1016/j.epsl.2013.01.036
Morishima, R., Stadel, J., Moore, B., 2010. From planetesimals to terrestrial planets: N-body simulations including the effects of nebular gas and giant planets. Icarus 207, 517–535. https://doi.org/10.1016/j.icarus.2009.11.038
Murray, C.D., Dermott, S.F., 1999. Solar system dynamics. Sol. Syst. Dyn. CD Murray SF McDermott Camb. UK Camb. Univ. Press ISBN 0-521-57295-9 Hc ISBN 0-521-57297-4 Pbk.
Nagasawa, M., Ida, S., Tanaka, H., 2001. Origin of high orbital eccentricity and inclination of asteroids. Earth Planets Space 53, 1085–1091. https://doi.org/10.1186/BF03351707
Nagasawa, M., Lin, D.N.C., Thommes, E., 2005. Dynamical Shake-up of Planetary Systems. I. Embryo Trapping and Induced Collisions by the Sweeping Secular Resonance and Embryo-Disk Tidal Interaction. Astrophys. J. 635, 578. https://doi.org/10.1086/497386
Nagasawa, M., Tanaka, H., Ida, S., 2000. Orbital Evolution of Asteroids during Depletion of the Solar Nebula. Astron. J. 119, 1480. https://doi.org/10.1086/301246
Nimmo, F., O'Brien, D.P., Kleine, T., 2010. Tungsten isotopic evolution during late-stage accretion: Constraints on Earth–Moon equilibration. Earth Planet. Sci. Lett. 292, 363–370. https://doi.org/10.1016/j.epsl.2010.02.003
O'Brien, D.P., Morbidelli, A., Levison, H.F., 2006. Terrestrial planet formation with strong dynamical friction. Icarus 184, 39–58. https://doi.org/10.1016/j.icarus.2006.04.005
Odert, P., Lammer, H., Erkaev, N.V., Nikolaou, A., Lichtenegger, H.I.M., Johnstone, C.P., Kislyakova, K.G., Leitzinger, M., Tosi, N., 2018. Escape and fractionation of volatiles and noble gases from Mars-sized planetary embryos and growing protoplanets. Icarus 307, 327–346. https://doi.org/10.1016/j.icarus.2017.10.031
Ogihara, M., Ida, S., Morbidelli, A., 2007. Accretion of terrestrial planets from oligarchs in a turbulent disk. Icarus 188, 522–534. https://doi.org/10.1016/j.icarus.2006.12.006
Papaloizou, J.C.B., Larwood, J.D., 2000. On the orbital evolution and growth of protoplanets embedded in a gaseous disc. Mon. Not. R. Astron. Soc. 315, 823–833. https://doi.org/10.1046/j.1365-8711.2000.03466.x
Pollack, J.B., Hubickyj, O., Bodenheimer, P., Lissauer, J.J., Podolak, M., Greenzweig, Y., 1996. Formation of the Giant Planets by Concurrent Accretion of Solids and Gas. Icarus 124, 62–85. https://doi.org/10.1006/icar.1996.0190
Portegies Zwart, S., 2020. The ecological impact of high-performance computing in astrophysics. Nat. Astron. 4, 819–822. https://doi.org/10.1038/s41550-020-1208-y
Quintana, E.V., Barclay, T., Borucki, W.J., Rowe, J.F., Chambers, J.E., 2016. THE FREQUENCY OF GIANT IMPACTS ON EARTH-LIKE WORLDS. Astrophys. J. 821, 126. https://doi.org/10.3847/0004-637X/821/2/126
Raymond, S.N., Izidoro, A., 2017. The empty primordial asteroid belt. Sci. Adv. 3, e1701138. https://doi.org/10.1126/sciadv.1701138
Raymond, S.N., O'Brien, D.P., Morbidelli, A., Kaib, N.A., 2009. Building the terrestrial planets: Constrained accretion in the inner Solar System. Icarus 203, 644–662. https://doi.org/10.1016/j.icarus.2009.05.016
Raymond, S.N., Quinn, T., Lunine, J.I., 2006. High-resolution simulations of the final assembly of Earth-like planets I. Terrestrial accretion and dynamics. Icarus 183, 265–282. https://doi.org/10.1016/j.icarus.2006.03.011
Render, J., Fischer-Gödde, M., Burkhardt, C., Kleine, T., 2017. The cosmic molybdenum-neodymium isotope correlation and the building material of the Earth. Geochem. Perspect. Lett. 170–178. https://doi.org/10.7185/geochemlet.1720
Ribeiro, R. de S., Morbidelli, A., Raymond, S.N., Izidoro, A., Gomes, R., Vieira Neto, E., 2020. Dynamical evidence for an early giant planet instability. Icarus 339, 113605. https://doi.org/10.1016/j.icarus.2019.113605
Rudge, J.F., Kleine, T., Bourdon, B., 2010. Broad bounds on Earth's accretion and core formation constrained by geochemical models. Nat. Geosci. 3, 439–443. https://doi.org/10.1038/ngeo872
Samuel, H., 2012. A re-evaluation of metal diapir breakup and equilibration in terrestrial magma oceans. Earth Planet. Sci. Lett. 313–314, 105–114. https://doi.org/10.1016/j.epsl.2011.11.001
Sanloup, C., Jambon, A., Gillet, P., 1999. A simple chondritic model of Mars. Phys. Earth Planet. Inter. 112, 43–54. https://doi.org/10.1016/S0031-9201(98)00175-7
Schiller, M., Bizzarro, M., Fernandes, V.A., 2018. Isotopic evolution of the protoplanetary disk and the building blocks of Earth and the Moon. Nature 555, 507–510. https://doi.org/10.1038/nature25990
Schiller, M., Bizzarro, M., Siebert, J., 2020. Iron isotope evidence for very rapid accretion and differentiation of the proto-Earth. Sci. Adv. 6, eaay7604. https://doi.org/10.1126/sciadv.aay7604
Stadel, J.G., 2001. Cosmological N-body simulations and their analysis. PhD Thesis.
Stökl, A., Dorfi, E., Lammer, H., 2015. Hydrodynamic simulations of captured protoatmospheres around Earth-like planets. Astron. Astrophys. 576, A87. https://doi.org/10.1051/0004-6361/201423638
Stökl, A., Dorfi, E.A., Johnstone, C.P., Lammer, H., 2016. DYNAMICAL ACCRETION OF PRIMORDIAL ATMOSPHERES AROUND PLANETS WITH MASSES BETWEEN 0.1 AND 5M$øplus$IN THE HABITABLE ZONE. Astrophys. J. 825, 86. https://doi.org/10.3847/0004-637X/825/2/86
Strom, K.M., Strom, S.E., Edwards, S., Cabrit, S., Skrutskie, M.F., 1989. Circumstellar material associated with solar-type pre-main-sequence stars - A possible constraint on the timescale for planet building. Astron. J. 97, 1451–1470. https://doi.org/10.1086/115085

Tanaka, H., Takeuchi, T., Ward, W.R., 2002. Three-Dimensional Interaction between a Planet and an Isothermal Gaseous Disk. I. Corotation and Lindblad Torques and Planet Migration. Astrophys. J. 565, 1257. https://doi.org/10.1086/324713

Tanaka, H., Ward, W.R., 2004. Three-dimensional Interaction between a Planet and an Isothermal Gaseous Disk. II. Eccentricity Waves and Bending Waves. Astrophys. J. 602, 388. https://doi.org/10.1086/380992

Tang, H., Dauphas, N., 2014. 60Fe–60Ni chronology of core formation in Mars. Earth Planet. Sci. Lett. 390, 264–274. https://doi.org/10.1016/j.epsl.2014.01.005

Toplis, M.J., Mizzon, H., Monnereau, M., Forni, O., McSween, H.Y., Mittlefehldt, D.W., McCoy, T.J., Prettyman, T.H., Sanctis, M.C.D., Raymond, C.A., Russell, C.T., 2013. Chondritic models of 4 Vesta: Implications for geochemical and geophysical properties. Meteorit. Planet. Sci. 48, 2300–2315. https://doi.org/10.1111/maps.12195

Tsiganis, K., Gomes, R., Morbidelli, A., Levison, H.F., 2005. Origin of the orbital architecture of the giant planets of the Solar System. Nature 435, 459–461. https://doi.org/10.1038/nature03539

Tu, L., Johnstone, C.P., Güdel, M., Lammer, H., 2015. The extreme ultraviolet and X-ray Sun in Time: High-energy evolutionary tracks of a solar-like star. Astron. Astrophys. 577, L3. https://doi.org/10.1051/0004-6361/201526146

Vockenhuber, C., Oberli, F., Bichler, M., Ahmad, I., Quitté, G., Meier, M., Halliday, A.N., Lee, D.-C., Kutschera, W., Steier, P., Gehrke, R.J., Helmer, R.G., 2004. New Half-Life Measurement of $^{182}$Hf: Improved Chronometer for the Early Solar System. Phys. Rev. Lett. 93, 172501. https://doi.org/10.1103/PhysRevLett.93.172501

Vorobyov, E.I., Elbakyan, V.G., 2018. Gravitational fragmentation and formation of giant protoplanets on orbits of tens of au. Astron. Astrophys. 618, A7. https://doi.org/10.1051/0004-6361/201833226

Wallace, S.C., Quinn, T.R., 2019. N-body simulations of terrestrial planet growth with resonant dynamical friction. Mon. Not. R. Astron. Soc. 489, 2159–2176. https://doi.org/10.1093/mnras/stz2284

Walsh, K.J., Levison, H.F., 2019. Planetesimals to terrestrial planets: Collisional evolution amidst a dissipating gas disk. Icarus 329, 88–100. https://doi.org/10.1016/j.icarus.2019.03.031

Walsh, K.J., Morbidelli, A., Raymond, S.N., O'Brien, D.P., Mandell, A.M., 2011. A low mass for Mars from Jupiter's early gas-driven migration. Nature 475, 206–209. https://doi.org/10.1038/nature10201

Wang, H., Weiss, B.P., Bai, X.-N., Downey, B.G., Wang, Jun, Wang, Jiajun, Suavet, C., Fu, R.R., Zucolotto, M.E., 2017. Lifetime of the solar nebula constrained by meteorite paleomagnetism. Science 355, 623–627. https://doi.org/10.1126/science.aaf5043

Ward, W.R., 1981. Solar nebula dispersal and the stability of the planetary system: I. Scanning secular resonance theory. Icarus 47, 234–264. https://doi.org/10.1016/0019-1035(81)90169-X

Warren, P.H., 2011. Stable-isotopic anomalies and the accretionary assemblage of the Earth and Mars: A subordinate role for carbonaceous chondrites. Earth Planet. Sci. Lett. 311, 93–100. https://doi.org/10.1016/j.epsl.2011.08.047

Wetherill, G.W., 1980. Formation of the Terrestrial Planets. Annu. Rev. Astron. Astrophys. 18, 77–113. https://doi.org/10.1146/annurev.aa.18.090180.000453

Wisdom, J., Holman, M., 1991. Symplectic maps for the n-body problem. Astron. J. 102, 1528–1538. https://doi.org/10.1086/115978

Woo, J.M.Y., Brasser, R., Matsumura, S., Mojzsis, S.J., Ida, S., 2018. The curious case of Mars' formation. Astron. Astrophys. 617, A17. https://doi.org/10.1051/0004-6361/201833148

Yamakawa, A., Yamashita, K., Makishima, A., Nakamura, E., 2010. CHROMIUM ISOTOPE SYSTEMATICS OF ACHONDRITES: CHRONOLOGY AND ISOTOPIC HETEROGENEITY OF THE INNER SOLAR SYSTEM BODIES. Astrophys. J. 720, 150–154. https://doi.org/10.1088/0004-637X/720/1/150

Yin, Q., Jacobsen, S.B., Yamashita, K., Blichert-Toft, J., Télouk, P., Albarède, F., 2002. A short timescale for terrestrial planet formation from Hf–W chronometry of meteorites. Nature 418, 949–952. https://doi.org/10.1038/nature00995

Yu, G., Jacobsen, S.B., 2011. Fast accretion of the Earth with a late Moon-forming giant impact. Proc. Natl. Acad. Sci. 108, 17604–17609. https://doi.org/10.1073/pnas.1108544108

Zahnle, K.J., Kasting, J.F., 1986. Mass fractionation during transonic escape and implications for loss of water from Mars and Venus. Icarus 68, 462–480. https://doi.org/10.1016/0019-1035(86)90051-5

Zheng, X., Lin, D.N.C., Kouwenhoven, M.B.N., 2017. Planetesimal Clearing and Size-dependent Asteroid Retention by Secular Resonance Sweeping during the Depletion of the Solar Nebula. Astrophys. J. 836, 207. https://doi.org/10.3847/1538-4357/836/2/207